\documentclass[prl, amsmath, amssymb, twocolumn, showpacs]{revtex4}% Physical Review B

\usepackage{graphicx}% Include figure files
\usepackage[caption=false]{subfig}
\usepackage{color}

\newcommand{\ket}[1]{\left\vert #1 \right\rangle}
\newcommand{\bra}[1]{\left\langle #1 \right\vert}

\newcommand{\imag}{\mathrm{i}}
\newcommand{\rd}{\mathrm{d}}

\renewcommand{\Im}{\mathrm{Im}}

\renewcommand{\vec}[1]{\boldsymbol{#1}}
\newcommand{\sO}{\sigma_0}
\newcommand{\sx}{\sigma_x}
\newcommand{\sy}{\sigma_y}
\newcommand{\sz}{\sigma_z}
\newcommand{\pO}{\pi_0}
\newcommand{\px}{\pi_x}
\newcommand{\py}{\pi_y}
\newcommand{\pz}{\pi_z}
\newcommand{\tO}{\tau_0}
\newcommand{\tx}{\tau_x}
\newcommand{\ty}{\tau_y}
\newcommand{\tz}{\tau_z}
\newcommand{\dphase}{}%\e^{\imag \phi}}
\newcommand{\dphaseC}{}%\e^{-\imag \phi}}
\newcommand{\up}{\uparrow}
\newcommand{\down}{\downarrow}
\newcommand{\sign}{\mathrm{sign}}

\newcommand{\HBdG}{H_{\mathrm{BdG}}}
\newcommand{\Hdis}{H_{\mathrm{dis}}}
\newcommand{\dsc}{d_{\mathrm{SC}}}

\begin{document}
\date{\today}

%opening
\title{Superconducting quantum spin-Hall systems with giant orbital g-factors} 
\author{R. W. Reinthaler, G. Tkachov, and E. M. Hankiewicz}
\affiliation{
Institute for Theoretical Physics and Astrophysics, W\"urzburg University, Am Hubland, 97074 W\"urzburg, Germany}

\pacs{	72.25.Dc, %Spin polarized transport in semiconductors
		73.23.Ad, %Ballistic Transport
		74.45.+c  %Proximity effect, Andreev reflection, SN and SNS
	 }

\begin{abstract}
Topological aspects of superconductivity in quantum spin-Hall systems (QSHSs) such as thin layers of three-dimensional topological insulators (3D Tis) or two-dimensional Tis 
are in the focus of current research. 
We examine hybrid QSHS/superconductor structures in an external magnetic field and predict a gapless superconducting state with protected edge modes.     
It originates entirely from the orbital magnetic-field effect caused by the locking of the electron spin to the momentum of the superconducting condensate flow. 
We show that such spin-momentum locking can generate a giant orbital g-factor of order of several hundreds, 
allowing one to achieve significant spin polarization in the QSHS in the fields well below the critical field of the superconducting material. 
We propose a three-terminal setup in which the spin-polarized edge superconductivity can be probed by Andreev reflection, leading to unusual transport characteristics: 
a non-monotonic excess current and a zero-bias conductance splitting in the absence of the Zeeman interaction.
\end{abstract}
\maketitle

%%%%%%%%%%%%%%%%%%%%%%%%%%%%%%%%%%%%%%%%%%%%%%%%%%%%%%%%%%%%%%%%%%%%%%%%
%%				Introduction											%%
%%%%%%%%%%%%%%%%%%%%%%%%%%%%%%%%%%%%%%%%%%%%%%%%%%%%%%%%%%%%%%%%%%%%%%%%
\textit{Introduction.} 
Spin-Hall effects are one of the most active fields in modern solid state physics \cite{Dyakonov1971,Murakami2003, Sinova2004, Hankiewicz2009, Kane2005, Koenig2007Science, qi_2011, tkachov_2013c}. 
In particular, the quantum spin-Hall effect \cite{Kane2005, Bernevig2006, Koenig2007Science, Knez2011} 
allows one to generate and convert charge and spin currents in protected edge channels \cite{Bruene2012}.  
Combining quantum spin-Hall systems (QSHSs) with superconductors (SCs) 
leads to a broader spectrum of interesting observable phenomena \cite{Knez2012, Hart2014, Pribiag2014}. 
These include quantum interference effects reported in \cite{Hart2014, Pribiag2014}, indicating superconducting transport through the edge states in the QSH regime. 
Understanding edge superconductivity in QSHS/SC hybrids is also instrumental to the proposals to realize Majorana zero modes 
in topological insulators (see, e.g., \cite{Fu2009} and reviews \cite{Alicea2012,Beenakker2013}).     

In this paper we predict a unique magnetic-field response of QSHS/SC hybrids which is characterized by very large effective g-factors reaching the order of several hundreds. 
It originates from the locking of the electron spin to the momentum of the superconducting condensate flow generated by an external magnetic field.
We show that this orbital effect has the form similar to the Zeeman spin splitting in thin superconducting films \cite{Meservey1970}, but involves 
an effective g-factor determined by the parameters of the QSHS/SC structure, viz.: 
the edge-state velocity, $v$, the thickness of the SC material, $d_\mathrm{SC}$, and the London penetration depth, $\lambda_L$:
\begin{align}
	\label{eq:Gfactor}
	g_* = \frac{2 m v}{\hbar} \lambda_L\tanh \frac{d_\mathrm{SC}}{2\lambda_L}, 
\end{align}
where $m$ is the electron rest mass. 
Equation (\ref{eq:Gfactor}) suggests a simple way to engineer $g_*$ through an appropriate material and structure choice.
For typical lateral HgTe- or Bi$_2$(Te,Se)$_3$-based/Nb systems we predict $g_* \approx 200 \div 400$. 
Such giant effective g-factors allow one to control the proximity-induced edge superconductivity by weak magnetic fields that do not destroy the SC order parameter  
or violate the protection of the edge states against backscattering. In the following, we illustrate these points in detail by developing a microscopic theory for
the orbital spin splitting of the superconducting edge states and spin-dependent Andreev reflection.

\begin{figure}[b]
	\begin{center}
		\includegraphics[width=0.75\linewidth]{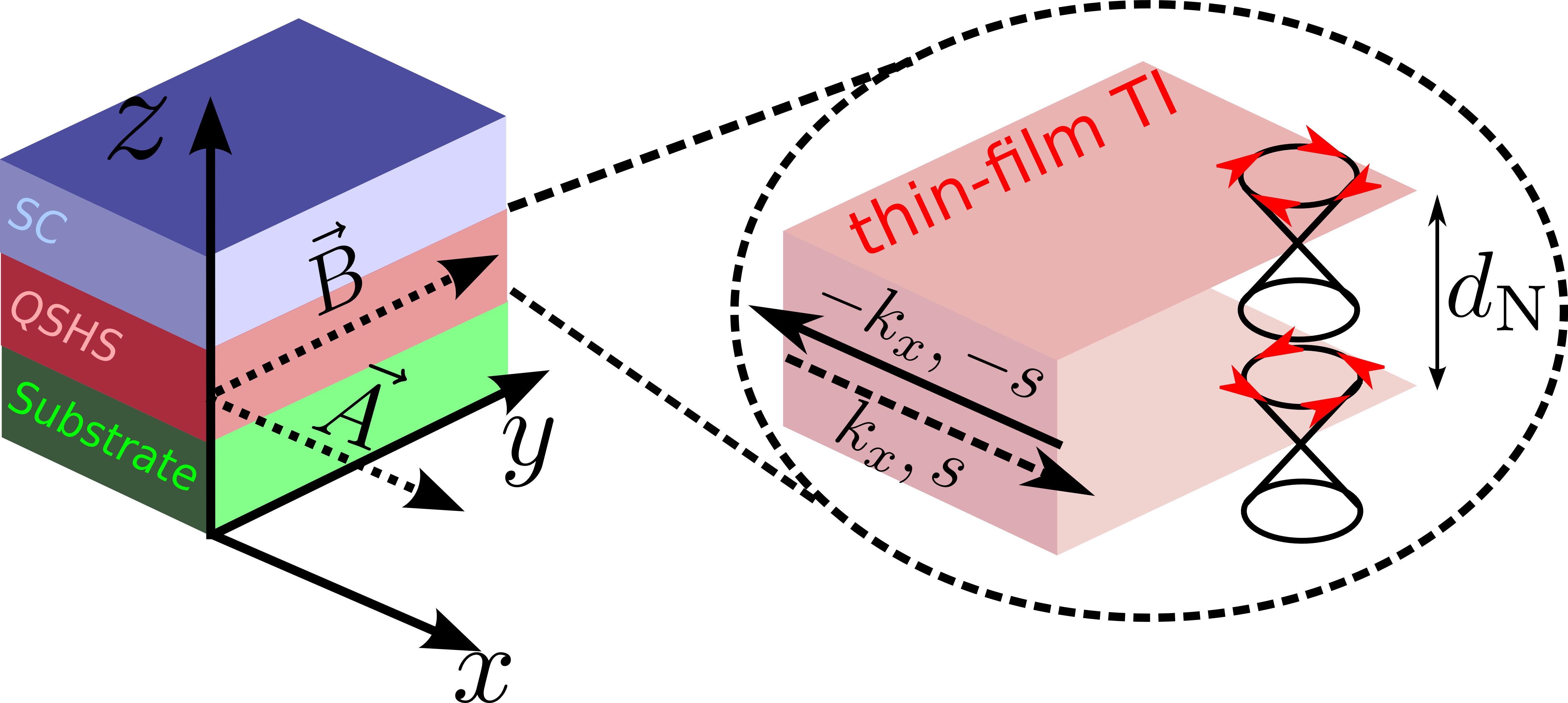}
		\caption{Schematic of a QSHS/SC structure on a substrate.  
		The system occupies semispace $-\infty<x< \infty, 0<y<\infty$. 
		Inset: QSHS realized in a thin-film topological insulator with hybridized top and bottom surfaces (see also text).}
		\label{fig:setup}
	\end{center}
\end{figure}

%%%%%%%%%%%%%%%%%%%%%%%%%%%%%%%%%%%%%%%%%%%%%%%%%%%%%%%%%%%%%%%%%%%%%%%%
%%				Model												%%
%%%%%%%%%%%%%%%%%%%%%%%%%%%%%%%%%%%%%%%%%%%%%%%%%%%%%%%%%%%%%%%%%%%%%%%%
\textit{Model.} 
We consider a lateral QSHS/SC structure (see Fig.~\ref{fig:setup}) with proximity-induced superconductivity in the QSHS described  
by the Bogoliubov-de Gennes (BdG) Hamiltonian
\begin{align}
\label{eq:Ham:BdG}
	\HBdG= 	\begin{pmatrix} 
				H_\mathrm{N} & \imag \dphase \Delta \tO\sy \\
				-\imag \dphaseC \Delta \tO\sy & -H^*_\mathrm{N}\\
			\end{pmatrix},
\end{align}
where $\Delta$ is the induced pairing potential, and $H_\mathrm{N}$ is the Hamiltonian of the normal QSHS.  
For the sake of concreteness, we assume that the QSHS is realized in a thin film of a 3D Ti \cite{Lu2010} 
and described by the four-band Hamiltonian \cite{Hamiltonian}: 
\begin{eqnarray}
H_\mathrm{N} =v (p_x + e A_x)\tz \sx +v p_y \tz\sy - \mu \tO\sO + \Gamma \tx \sO. 
\label{eq:Ham:3DTI}
\end{eqnarray}
It models top and bottom surface Dirac states hybridized in a thin slab with the coupling energy $\Gamma$. 
We use the representation in which the hybridization term is proportional to the off-diagonal Pauli matrix $\tx$ acting
in the basis of top and bottom states, while the Dirac terms are diagonal in this basis and proportional to the Pauli matrix $\tz$. 
Pauli matrices, $\sx$ and $\sy$, represent the spin of the surface carrier ($\sO$ is the corresponding unit matrix),  
$\vec{p}=-\imag\hbar [\partial_x, \partial_y, 0]$ is its momentum operator, and $\mu$ is the chemical potential. 
Additionally, we introduce an external magnetic field $\vec{B}=[0,B_y,0]$ and the vector potential $\vec{A}=[A_x(z),0,0]$ related by $B_y = \partial_z A_x(z)$ 
and choose $e>0$. The function $A_x(z)$ must be determined by taking into account the screening of the external field by the diamagnetic (London) current in the SC \cite{appendix}. 
Since realistically the thickness of the QSHS, $d_\mathrm{N}$, is much smaller than $d_\mathrm{SC}$, 
the vector potential in the QSHS can be approximated by the $A_x$ value taken at the SC surface 
$A_x \approx -B_y \lambda_L\tanh( d_\mathrm{SC}/2\lambda_L) $ \cite{appendix}.
In this approximation, the effect of the magnetic field is equivalent to a linearly varying superconducting phase, $\phi(x) = (2e/\hbar)A_x\, x$, 
whose gradient generates a condensate flow with the momentum $2eA_x$.

The relevant symmetry operations for our system are time reversal (TR) $\hat{\tau}$, particle-hole (PH) conjugation $\hat{C}$, and their combination $\hat{\tau}\hat{C}$. 
Introducing Pauli matrices in particle-hole space $\pi_{x,y,z}$ and the complex conjugation operation $\hat{K}$, we can define the symmetry operations and their action as
\begin{align}
\label{eq:TR:Op}
 &\hat{\tau} = - \imag \pO \tO \sy \hat{K},\; \HBdG (\vec{A}) \stackrel{\hat{\tau}}{\longrightarrow} \HBdG(-\vec{A}),\\
\label{eq:PH:Op}
 &\hat{C} = -  \px \tO \sO \hat{K},\; \HBdG (\vec{A})\stackrel{\hat{C}}{\longrightarrow} -\HBdG(\vec{A}), \\
\label{eq:TC:Op}
 &\qquad\qquad\qquad\quad\,\, \HBdG (\vec{A})\stackrel{ \hat{\tau}\hat{C} }{\longrightarrow} -\HBdG(-\vec{A}).
\end{align}
For $\vec{A}\not= 0$ the TR is broken, while the PH symmetry is preserved, so that the excitation spectrum remains symmetric around the Fermi level.

%%%%%%%%%%%%%%%%%%%%%%%%%%%%%%%%%%%%%%%%%%%%%%%%%%%%%%%%%%%%%%%%%%%%%%%%
%%				orbital spin polarization							%%
%%%%%%%%%%%%%%%%%%%%%%%%%%%%%%%%%%%%%%%%%%%%%%%%%%%%%%%%%%%%%%%%%%%%%%%%
\textit{Orbital spin splitting.} 
We seek the solution to the BdG equation $\HBdG \, \psi(x,y) = E \, \psi(x,y)$ in the form of a plane wave 
$
\psi(x,y) \propto e^{ \imag k x - y/\delta } 
$
propagating along $x$ with a wave vector $k$ 
and decaying exponentially for $y>0$ on a scale $\delta$. 
From the requirement that the normal component of the current vanishes at the edge, $j_y(x,y=0)=0$, we derive the boundary condition
\begin{align}
\label{eq:BC}
 \psi(x,y=0) = \pO \ty \sy \psi(x,y=0).
\end{align}
It is specific to Dirac fermions and preserves TR. In \cite{appendix} we prove that
the edge solution exists only for an inverted band gap $\Gamma<0$, with $\delta  = \hbar v/|\Gamma|$, 
and has the following structure
\begin{align}
	\label{eq:sol:spinor:BC}
	\psi(x,y) = \left(\chi_s , \imag \sy \chi_s, \xi_s , \imag \sy \xi_s \right)^T \, e^{ \imag k x - y/\delta }.
\end{align}
We took advantage of the one-dimensional character of the edge spin-momentum locking and used the eigenstates $\chi_s=(s, 1)^T$ of the spin matrix $\sigma_x$ 
with the eigenvalues $s=\pm 1$ as well as the related hole spinors $\xi_s$ (see \cite{appendix}). The edge dispersion is
\begin{align}
	\label{eq:dispersion}
	E^s_\pm (k) &= s v e A_x \pm \sqrt{\Delta^2 + (\mu - s \hbar v k)^2}\nonumber\\
      			   &= -s g_* \mu_{\mathrm{B}} B_y  \pm \sqrt{\Delta^2 + (\mu - s \hbar v k)^2}.
\end{align}
We note that the spin-momentum locking couples the spin $s$ to the condensate momentum $\propto A_x$,
giving rise to the orbital spin splitting of the levels with the effective g-factor $g_* $ introduced earlier in Eq. (\ref{eq:Gfactor}) 
[$\mu_{\mathrm{B}}$ is the Bohr magneton]. Let us estimate $g_*$ for Bi$_2$X$_3$ (where X=Se, Te) and HgTe-based structures which typically have $\hbar v \approx 300\;\mathrm{meV\; nm}$ \cite{Liu2010b,Bruene2011} 
and, consequently, $g_* \approx 7.9 \;\mathrm{nm^{-1}} \times \lambda_L\tanh(d_\mathrm{SC}/2\lambda_L)$. 
For commonly used SC Nb with $\lambda_L \approx 50$ nm \cite{Maxfield1965} and nanoscale thicknesses $d_\mathrm{SC}= 50 \div 250$ nm, 
we obtain $g_* \approx 200 \div 400$, which exceeds the electron spin g-factor by at least two orders of magnitude. 
\begin{figure}[tb]
	\begin{center}
		\subfloat[]{\includegraphics[width=0.5\linewidth]{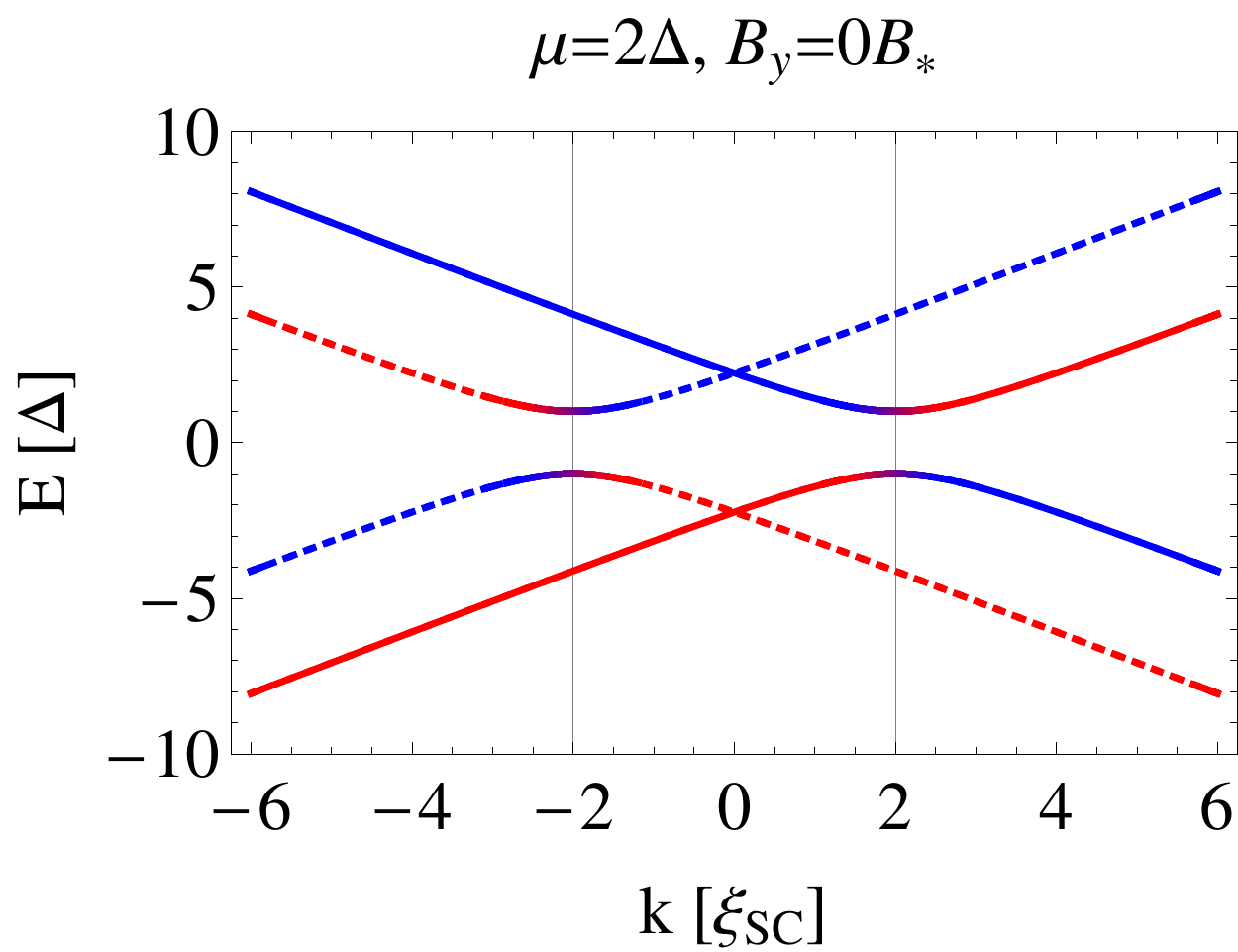}\label{fig:dispersions:a}}\hfill
		\subfloat[]{\includegraphics[width=0.5\linewidth]{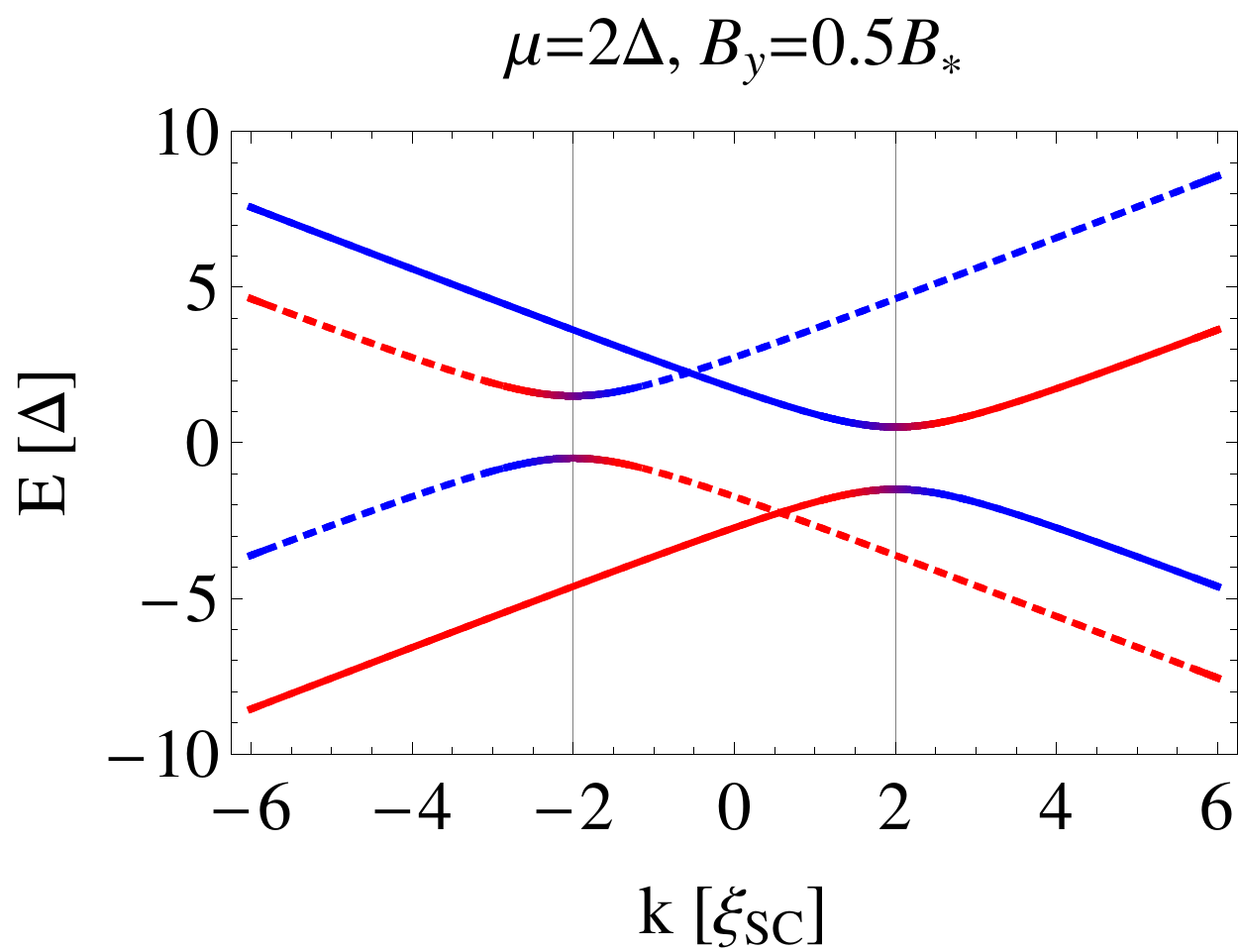}\label{fig:dispersions:b}}\\
		\subfloat[]{\includegraphics[width=0.5\linewidth]{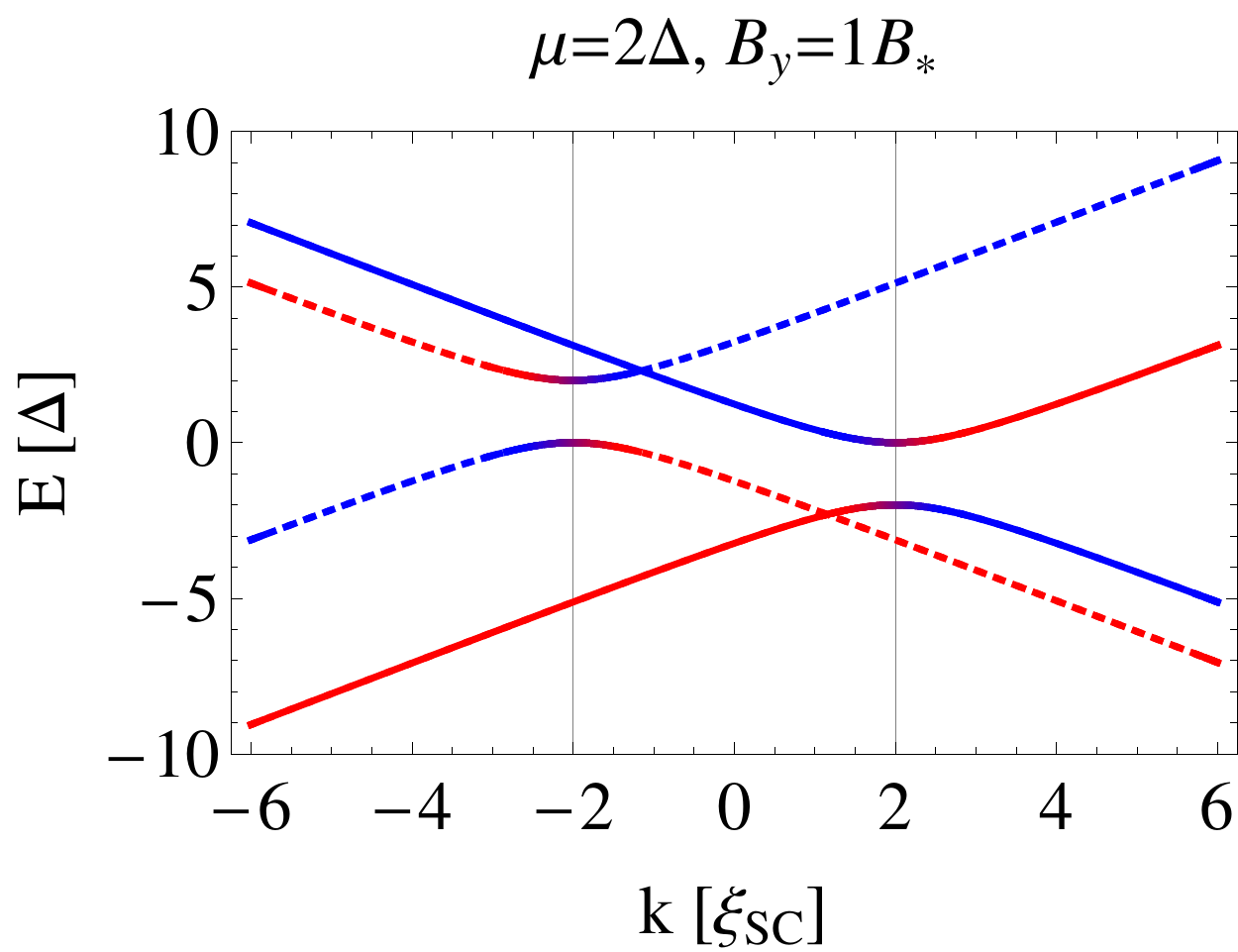}\label{fig:dispersions:c}}\hfill
		\subfloat[]{\includegraphics[width=0.5\linewidth]{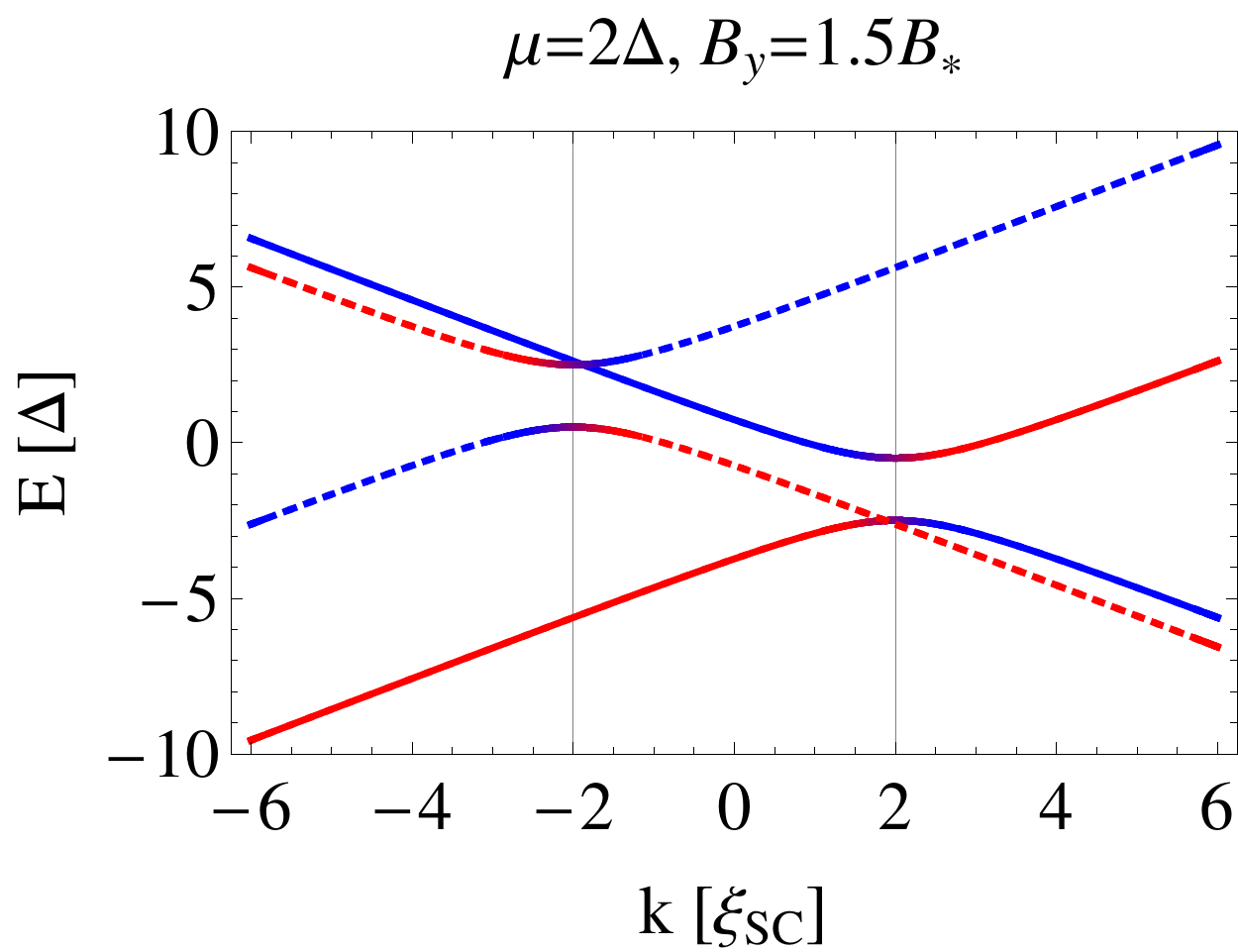}\label{fig:dispersions:d}}\\
		\caption{Edge dispersion for different parameters. Solid and dashed curves correspond to the opposite-spin states with $s=\pm 1$, respectively. 
		Red and blue colors schematically indicate particle and hole branches corresponding to those in the normal system. For $B_y=0$ there is an excitation gap at two Fermi points
		$\pm k_F$ indicated by the vertical lines. For $B_y\neq0$ the gap is reduced at each Fermi point (b), vanishing completely for $B_y= B_*$ (c). 
		For $B_y > B_*$ (d) the spectrum is gapless, albeit each spin branch separately has a gap at finite energies. }
		\label{fig:dispersions}
	\end{center}
\end{figure}
In contrast to the Zeeman spin splitting \cite{Meservey1970}, the orbital effect considered here does not require high magnetic fields to generate spin splitting, as will be shown below.
At $\mu=A_x=\Delta=0$ Eq.~\eqref{eq:dispersion} describes a spin-degenerated Dirac cone. A finite chemical potential shifts the electron and hole branches relative to each other, 
so that the SC pairing opens a gap at two Fermi points $\pm k_F = \pm \mu/(\hbar v)$, cf. Fig.~\ref{fig:dispersions:a} for $A_x=B_y=0$. 
Note that the Fermi points $\pm k_F$ correspond to the opposite-spin states with $s=\pm 1$. 
Therefore, the magnetic field splits the states at different Fermi points, shifting them relative to each other in energy, as depicted in Fig.~\ref{fig:dispersions:b}.
When the orbital energy scale matches the induced gap at $\vert A_x\vert  = \Delta/(e v)$ or, equivalently $\vert B_y\vert  =B_*$, with 
\begin{align}
\label{eq:GapClosing}
	B_*= \frac{\Delta}{e v \lambda_L\tanh(d_\mathrm{SC}/2\lambda_L)}= \frac{\hbar}{e \xi_{\mathrm{SC}}\lambda_L\tanh(d_\mathrm{SC}/2\lambda_L)},
\end{align}
where $\xi_{\mathrm{SC}}$ is the coherence length, the excitation gap vanishes [Fig.~\ref{fig:dispersions:c}]. For the induced gap $\Delta \lessapprox 0.1\;\mathrm{meV}$ \cite{Maier2012} and Nb-structure parameters used above, 
$B_* \lessapprox 5.5 \;\mathrm{mT}$ which is much smaller than the critical field of Nb. 
For $B_y > B_*$ the spectrum remains gapless. Note, however, that each spin branch separately has a gap at finite energies [Fig.~\ref{fig:dispersions:d}], 
which is the origin of spin-dependent Andreev reflection discussed later.
%

%%%%%%%%%%%%%%%%%%%%%%%%%%%%%%%%%%%%%%%%%%%%%%%%%%%%%%%%%%%%%%%%%%%%%%%%
%%				Symmetries and protection							%%
%%%%%%%%%%%%%%%%%%%%%%%%%%%%%%%%%%%%%%%%%%%%%%%%%%%%%%%%%%%%%%%%%%%%%%%%
\textit{Symmetries and protection.} 
Let us examine the properties of the edge solutions at fixed energy $E$. 
For that purpose, we solve \eqref{eq:dispersion} for the wave vector
\begin{eqnarray}
	\label{eq:wavevector}
		k^\beta_{Es} = s k_F + \beta \sqrt{(E- s v e A_x)^2 - \Delta^2 }/\hbar v, \; \beta= \pm1,\quad
\end{eqnarray}
and find the energy-dependent eigenfunctions as
\begin{align}
	\label{eq:sol:Psi:Es}
	\psi^\beta_{Es}(y) = 
    \frac{
	\exp[\imag k^\beta_{Es} x - y/\delta]
    }
	{\sqrt{2\delta}}
	\begin{pmatrix}
		u^\beta_{Es} \phi_s \\
		v^\beta_{Es} \phi_{-s} \\	
	\end{pmatrix}.
\end{align}
It is expressed in terms of the normal-state edge solution $\phi_s =  (s , 1, 1, -s)^T$ and the BCS coherence factors
\begin{eqnarray}
	\label{eq:sol:uv:Es}
	u^{\beta}_{Es} &=&\frac{1}{\sqrt{2}}\frac{\Delta}{ \sqrt{D_{Es}^2 - s \beta D_{Es}\sqrt{D_{Es}^2 - \Delta^2}}}, \nonumber\\
	v^{\beta}_{Es} &=& s \; \sign[D_{Es}] u_{Es}^{-\beta }, \quad D_{Es} = E - s v e A_x.
\end{eqnarray}
Using $k^\beta_{Es} (A_x)= - k^{-\beta}_{E-s} (-A_x)$, $u^\beta_{Es} (A_x) = u^{-\beta}_{E-s}(-A_x)$ as well as 
$v^\beta_{Es} (A_x) = -v^{-\beta}_{E-s}(-A_x)$ the action of TR on the eigenstates is described by
\begin{align}
	\label{eq:TR:psi}
	\psi^\beta_{Es} (A_x) \stackrel{\hat{\tau}}{\longrightarrow} s \psi^{-\beta}_{E-s} (-A_x).
\end{align}
For $A_x=0$ the states $\psi^{\beta}_{Es}$ and $\hat\tau \psi^\beta_{Es}$ are Kramers' partners. 
Using \eqref{eq:sol:uv:Es} the corresponding relations for PH and $\hat{\tau}\hat{C}$ are derived:
\begin{align}
	\label{eq:PH:psi}
	\psi^{\beta}_{Es} &\stackrel{\hat{C}}{\longrightarrow}- s \;\sign D_{Es} \psi^{-\beta}_{-E-s}\\
	\label{eq:TC:psi}
	\psi_{Es}^\beta &\stackrel{\hat{ \tau}\hat{C} }{\longrightarrow} \sign D_{Es} \exp\left[ \imag \frac{2 \beta \sqrt{D_{Es}^2 - 
	\Delta^2} }{\hbar v} x  \right] \pz \tO \sO \psi^{-\beta}_{Es}.
\end{align}
In contrast to \eqref{eq:TR:psi} and \eqref{eq:PH:psi} the relation for $\hat{\tau}\hat{C}$ is not broken by a finite vector potential and connects states at the same energy. 
The price is an $x$-dependent phase factor and a transformation in particle-hole space. 
%Comparison with \eqref{eq:sol:QPC:ES} shows that 
$\hat{C}$ and $\hat{\tau}\hat{C}$ both change the quasi-particle character.\\
In the following we discuss the protection of the superconducting edge states against elastic scattering, 
assuming a non-magnetic disorder potential $\Hdis = V(x) \pz \tO \sO$. 
To that end, we calculate the matrix element of $\Hdis$ between states with the same energy $E$:  %	O^{\beta_1\beta_2}_{E s_1s_2} :=& 
\begin{align}
	\label{eq:OV}
	\bra{\psi^{\beta_1}_{Es_1}} \Hdis \ket{\psi^{\beta_2}_{Es_2}} 
		&= \frac{V(x)}{2} \exp\left[\imag (k^{\beta_2}_{Es_2} - k^{\beta_1}_{Es_1})x\right]   \nonumber \\
		\cdot (1+ s_1s_2) & \left(\left(u^{\beta_1}_{Es_1}\right)^* u^{\beta_2}_{Es_2}  - \left(v^{\beta_1}_{Es_1}\right)^* v^{\beta_2}_{Es_2} \right).
\end{align}
The matrix element vanishes for the states with opposite spin projections $s_1=-s_2$. 
This means the absence of backscattering that couples different Fermi points $k^\pm_F=\pm \mu/\hbar v$. 
This is true irrespective of the presence or absence of the magnetic field.  
Furthermore, there is no scattering between the states near each Fermi point, $k^+_F$ or $k^-_F$.   
For such states $s_1=s_2=s$, $\beta_1 = - \beta_2=\beta$, and $(u^{\beta}_{Es})^* u^{-\beta}_{Es} = (v^{\beta}_{Es})^* v^{-\beta}_{Es}$, 
also yielding vanishing matrix element (\ref{eq:OV}). 
This can be interpreted as protection by the $\hat{\tau}\hat{C}$ symmetry, which has the same matrix structure as $\Hdis$, leading to a generalized Kramers' theorem 
\begin{align}
	\label{eq:TC:KramersTheorem}
 \bra{\psi^\beta_{Es} } \hat{\tau}\hat{C} \ket{\psi^\beta_{Es}} \propto  \bra{\psi^\beta_{Es} } \pz\tO\sO \ket{\psi^{-\beta}_{Es}}=0.
\end{align}
Physically, the $\hat{\tau}\hat{C}$ protection means that the states with opposite $\beta$s (i.e. particles and holes) cannot be converted into each other without 
being Andreev reflected \cite{Blonder1982}. For $B_y > B_*$ \eqref{eq:GapClosing} there are four protected zero energy states.
At the transition $B_y=B_*$ their number is reduced to two:
\begin{align}
	\label{eq:sol:zeroE}
	\psi^0_s (y) = 
	\frac{
	\exp[\imag s k_F x - y/\delta]
    }
	{\sqrt{2\delta}}
	\begin{pmatrix} \phi_s \\ - \sign A_x \phi_{-s} \end{pmatrix}.
\end{align}
These are equivalent to two Majorana zero modes. Indeed, defining $\eta = (1 + \sign A_x)/2$, one can construct new states $\gamma_1 = \exp[\imag \eta \pi/2] (\psi^0_s + \psi^0_{-s})$ 
and $\gamma_2 = -\imag  \exp[\imag \eta \pi/2] (\psi^0_s - \psi^0_{-s})$ with self-adjoint properties of emergent Majorana fermions.

%%%%%%%%%%%%%%%%%%%%%%%%%%%%%%%%%%%%%%%%%%%%%%%%%%%%%%%%%%%%%%%%%%%%%%%%
%%				Proposed experimental setup							%%
%%%%%%%%%%%%%%%%%%%%%%%%%%%%%%%%%%%%%%%%%%%%%%%%%%%%%%%%%%%%%%%%%%%%%%%%
\textit{Proposed experimental detection.} 
The spin splitting in the QSHS can be characterized by the spin-dependent density of states (sDOS) defined as
	$\rho_s ( E)= -\int_0^\infty \rd y \frac{1}{\pi} \mathrm{Tr} \;\left[\Im G^R_s (y,y)\right]_{11} $
	where $\left[\Im G^R_s (y,y)\right]_{11}$ is the imaginary part of the quasiparticle retarded Green's function \cite{appendix}.
$\rho_s(E)$ diverges at the band edges $E= s ev A_x \mp \Delta$ and approaches $1 /(\hbar v)$ away from the gap. 
\begin{figure}
	\begin{center}
		\subfloat[]{\includegraphics[width=0.49\linewidth]{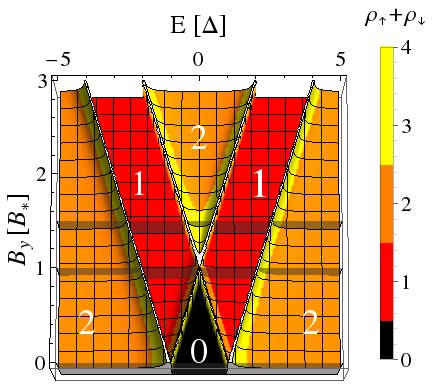}\label{fig:DOS}}\hfill
		\subfloat[]{\includegraphics[width=0.49\linewidth]{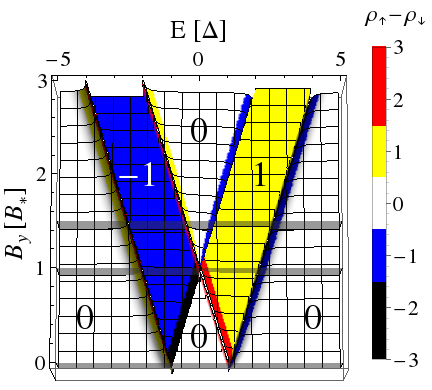}\label{fig:sDOS}}
		\caption{The DOS (a) and spectral spin polarization (b) as a function of $B_y$ and $E$. The gap closing is clearly visible at $B_y = B_*$. 
The shaded planes indicate the $B_y$ values used in Figs~\ref{fig:dispersions:a}, \ref{fig:dispersions:c} and \ref{fig:dispersions:d} respectively.}
		\label{fig:DOS:sDOS}
	\end{center}
\end{figure}
The sum $\rho_\up(E) + \rho_\down(E)$ yields the DOS, while the difference $\rho_\up(E) - \rho_\down(E)$ characterizes the spectral spin polarization. 
These quantities are plotted as a function of $E$ and $B_y$ in Fig.~\ref{fig:DOS:sDOS}. The gray shaded planes indicate $B_y/B_*=0$, $1$ and $1.5$ used in Fig.~\ref{fig:dispersions}. 
At zero magnetic field the DOS shows a quasi-particle gap between $E=\pm \Delta$. 
Within the bands the DOS is constant and spin degenerated. 
Hence the spin polarization vanishes. A finite $B_y$ shifts the energy of the s-branches by $-s g_{*}\mu_{\mathrm{B}} B_y$. 
This is reflected by a splitting of the peaks in the DOS. 
At gap closing \eqref{eq:GapClosing} the two peaks cross and the gaps for the different $s$-branches separate in energy. 
In the energy range, where only one $s$-branch is gaped, a finite spin polarization arises. 

\begin{figure}
	\begin{center}
		\subfloat[]{\includegraphics[width=0.5
	\linewidth]{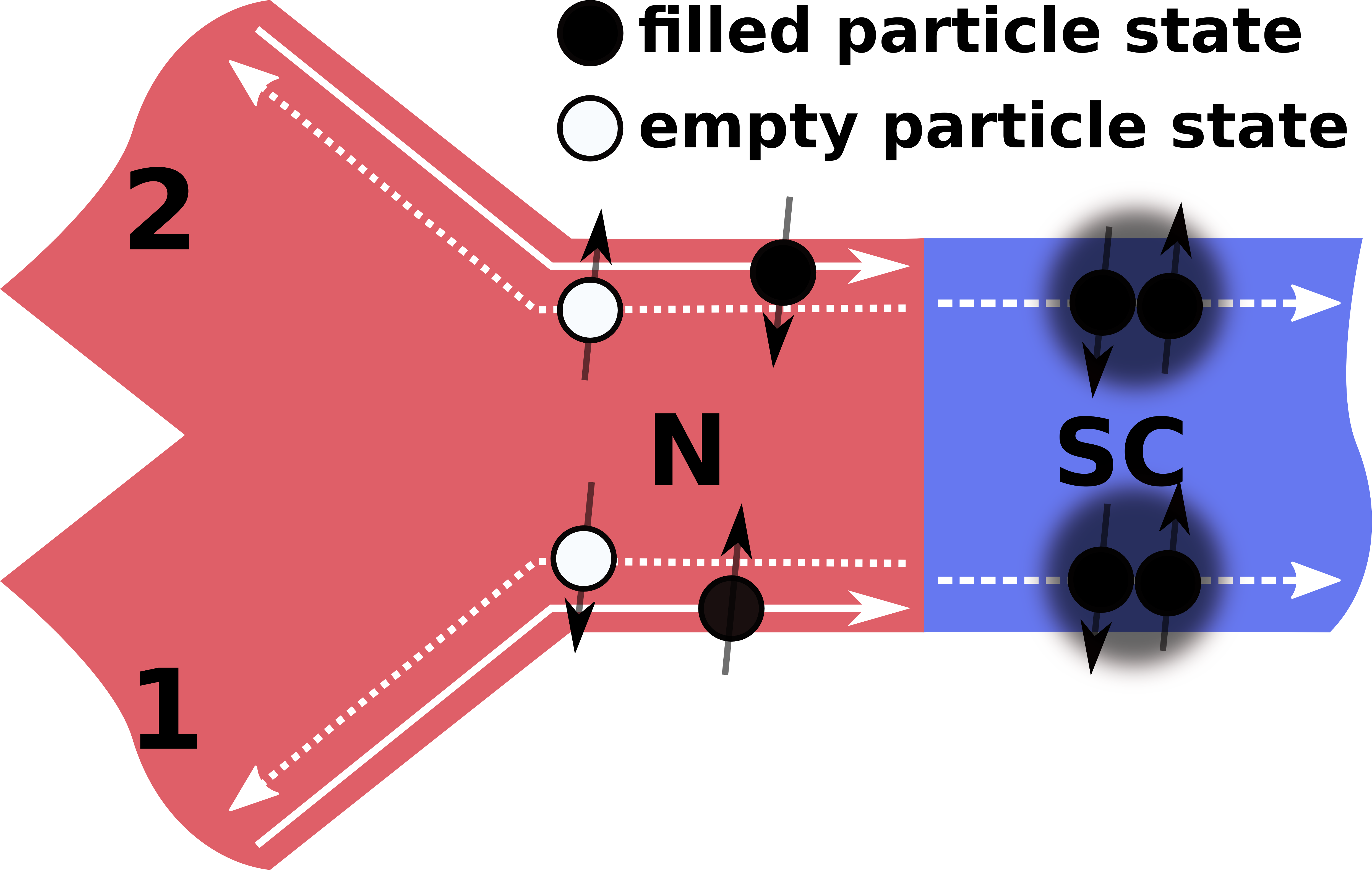}\label{fig:Transport:Setup}}\hfill
		\subfloat[]{\includegraphics[width=0.5\linewidth]{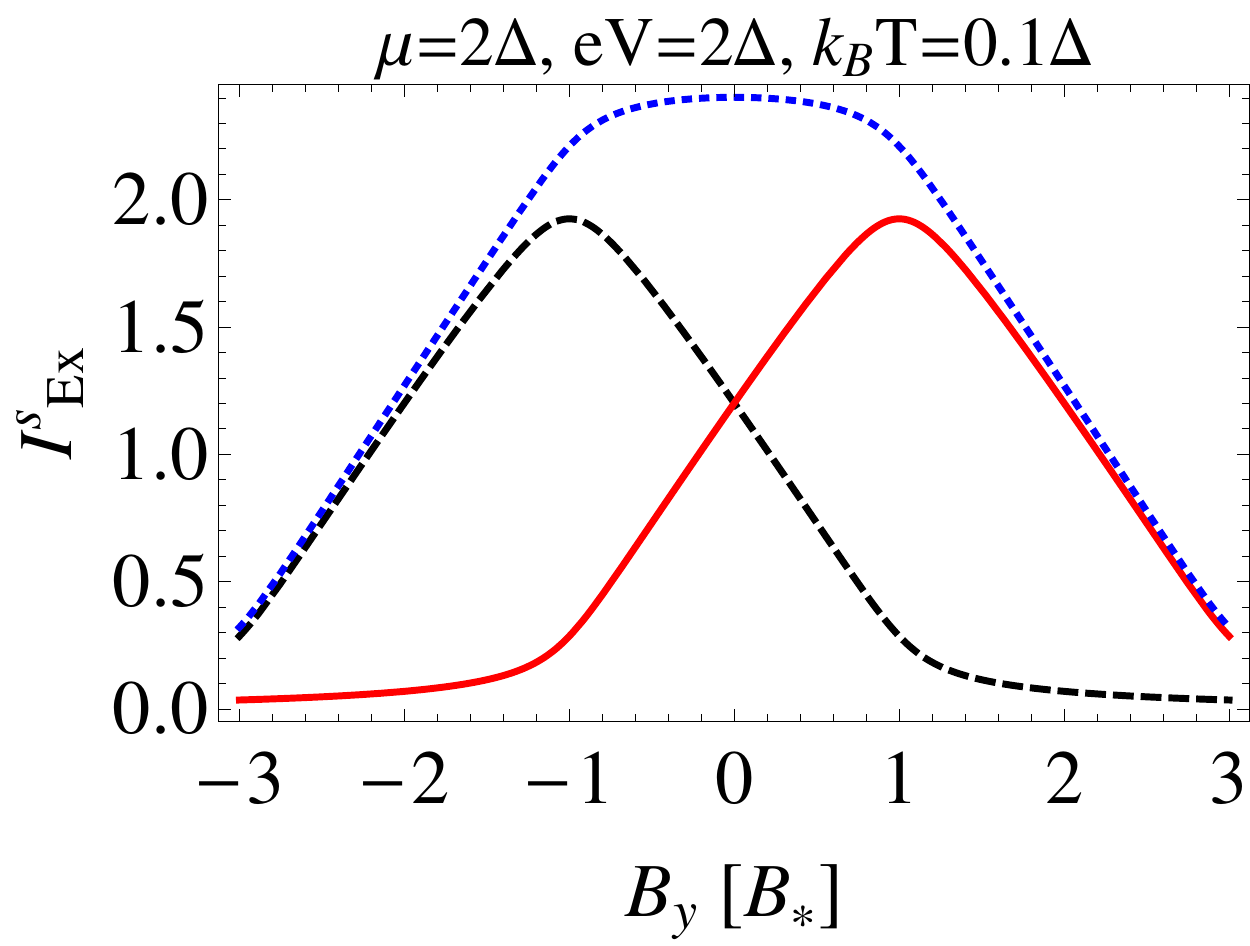}\label{fig:Transport:Iex}}\\
		\subfloat[]{\includegraphics[width=0.5
	\linewidth]{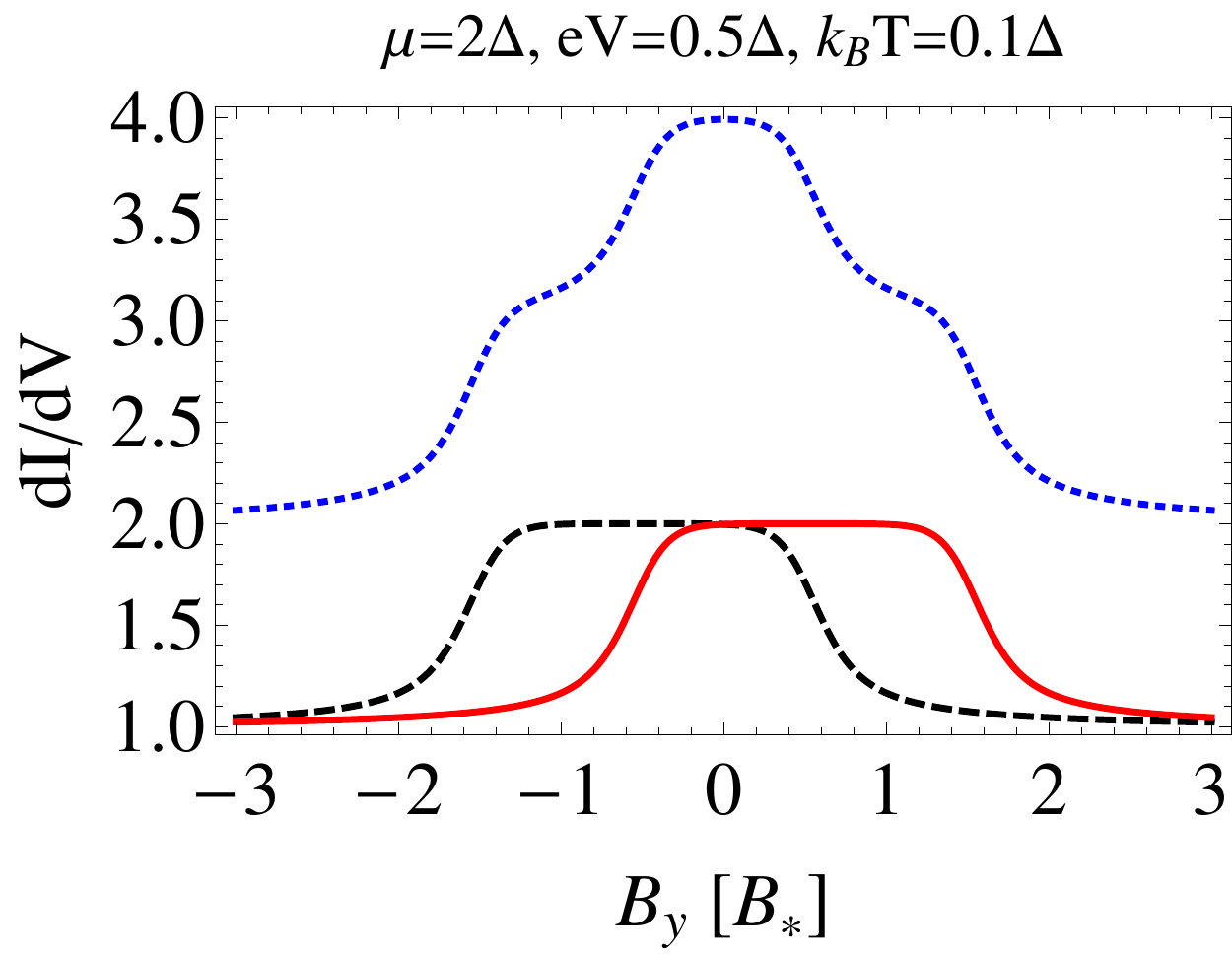}\label{fig:Transport:dIdV:a}}\hfill
		\subfloat[]{\includegraphics[width=0.5\linewidth]{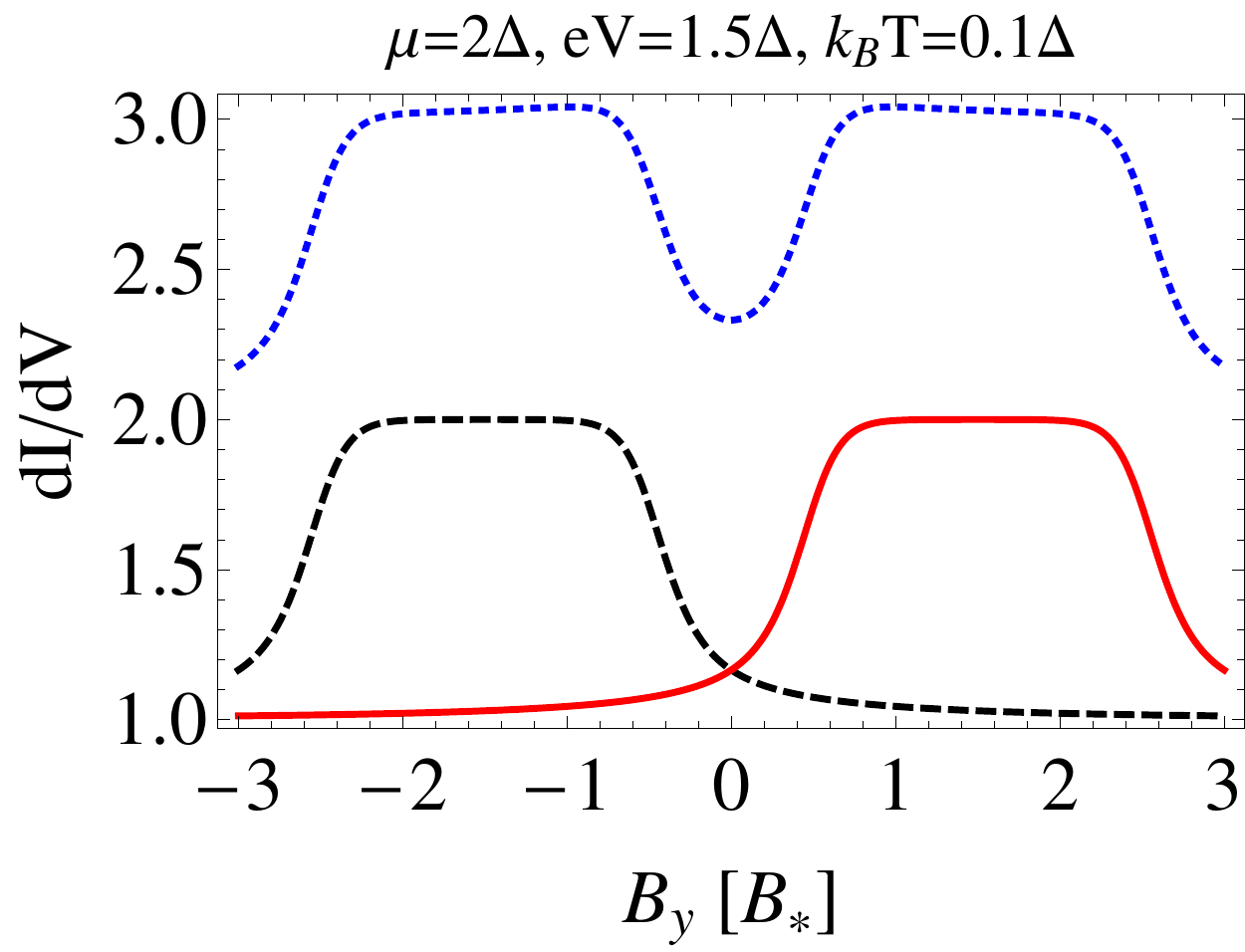}\label{fig:Transport:dIdV:b}}
		\caption{(a) Y-shaped junction between a normal (N) and a superconducting (SC) 3D Ti thin film. The different spin branches, depicted as black, dashed (s=1) and red, solid (s=-1) lines in (b),(c) and (d), can be measured independently in the leads 1 or 2 and summed up to the total signal (blue, dotted). The excess current $I_{\mathrm{ex}}$ (b) is odd in $B_y$ for a single spin, showing a clear maximum at finite $B_y$. The $\rd I/\rd V$ characteristics have a pronounced maximum at zero field, which is split when $eV > \Delta$ (d).}
		\label{fig:Transport}
	\end{center}
\end{figure}

Although the spin $s$, which characterizes the states, is derived from a $\sx$ eigenstate $\chi_s$, 
the structure of the solution \eqref{eq:sol:spinor:BC} always combines parts with opposite $\sx$ expectation values. 
The reason is that the eigenstates of \eqref{eq:Ham:3DTI} consist of a perfect mixture of the two surface states with opposite helicity. 
Hence it is to be understood as a pseudo-spin and the spin polarization does not result in a spin signal, 
which is directly measurable by a magnetic tip, like it was found for HgTe quantum wells \cite{Bruene2012}. 
However, comparing Figs.~\ref{fig:DOS} and \ref{fig:sDOS} we see that the sDOS is identically mapped on the regions with unit DOS, 
being therefore an indirect measure of the spin polarization.
A direct measurement of the spin polarization can be realized in transport experiments. 
We propose the Y-forked 3 terminal normal-superconducting (NSC) junction, sketched in Fig.~\ref{fig:Transport:Setup}. 
If the system is large enough, the counter-propagating edge states with opposite spin do not hybridize and can be treated independently, while  perfect Andreev reflection occurs  at the NSC interface \cite{Adroguer2010,Reinthaler2013}. 
When a bias $V$ is applied across the NSC junction we calculate the excess current $I_{\mathrm{ex}}$ as well as the $\rd I/\rd V$ characteristics within the BTK formalism \cite{Blonder1982, appendix} 
at finite temperature $k_{\mathrm{B}} T = 0.1\Delta$. Since the spin orientations are in one-to-one correspondence to the geometrical edges, 
the two spin branches can be detected directly by measuring contacts 1 and 2 independently. 
In Fig.~\ref{fig:Transport:Iex} we plot $I_{\mathrm{ex}}$ as a function of $B_y$. 
While the total current (blue dotted line) is symmetric, the individual spin branches are odd and show a distinct, spin-dependent maximum at 
$B_y = -s e V/(2g_{*}\mu_{\mathrm{B}})$ \cite{appendix}. 
Further we show the $\rd I/\rd V$ characteristics as a function of $B_y$ in Figs~\ref{fig:Transport:dIdV:a} and \ref{fig:Transport:dIdV:b}. 
The spin-dependent quasi-particle gaps correspond to the conductance plateaus and can be detected directly by measuring the non-local conductances in lead 1 and 2 separately. 
The peak in the total conductance at zero field splits when $eV$ exceeds the superconducting gap, i.e. when the quasi-particle gap is closed. 
This is a rather general signature of the spin-split superconducting states \cite{Tkachov2005}. 
Since energy and magnetic field enter the solutions \eqref{eq:sol:uv:Es} in the same way, 
an analogous signal can be obtained by varying $V$ for fixed $B_y$. 
The non-monotonic excess current and the split conductance peak are both hallmarks of the spin polarization in a superconducting QSHS.

%%%%%%%%%%%%%%%%%%%%%%%%%%%%%%%%%%%%%%%%%%%%%%%%%%%%%%%%%%%%%%%%%%%%%%%%
%%				Conclustion											%%
%%%%%%%%%%%%%%%%%%%%%%%%%%%%%%%%%%%%%%%%%%%%%%%%%%%%%%%%%%%%%%%%%%%%%%%%
In conclusion we have demonstrated the existence of a hybrid superconducting quantum spin-Hall system, which is robust against elastic backscattering in finite magnetic fields. 
With appropriately chosen structure and material parameters, the system is characterized by very large effective g-factors reaching the order of several hundreds,
allowing one to achieve spin polarization by applying a rather weak external magnetic field.
The helicity and the spin polarization of the emerging states can be experimentally detected in 3-terminal NSC-junctions. 
We predict a non-monotonic behavior of the excess current as well as splitting of  the zero field and the bias differential conductance peaks. 
Both features are connected with the closure of the quasi-particle gap at very small magnetic fields.

%%%%%%%%%%%%%%%%%%%%%%%%%%%%%%%%%%%%%%%%%%%%%%%%%%%%%%%%%%%%%%%%%%%%%%%%
%%				acknowledgements										%%
%%%%%%%%%%%%%%%%%%%%%%%%%%%%%%%%%%%%%%%%%%%%%%%%%%%%%%%%%%%%%%%%%%%%%%%%
We acknowledge fruitful discussions with Dietrich Rothe, Fran\c{c}ois Cr\'epin, and Teun Klapwijk.
We thank for the financial support the German Science Foundation (DFG), grants No HA 5893/4-1 within SPP 1666, HA5893/5-2 within
FOR1162 and TK60/1-1 (G.T.), as well the ENB graduate school "Topological insulators".

%%%%%%%%%%%%%%%%%%%%%%%%%%%%%%%%%%%%%%%%%%%%%%%%%%%%%%%%%%%%%%%%%%%%%%%%
%%				Bibliography											%%
%%%%%%%%%%%%%%%%%%%%%%%%%%%%%%%%%%%%%%%%%%%%%%%%%%%%%%%%%%%%%%%%%%%%%%%%
%\bibliography{SCspinHall}{}

\newpage

%%%%%%%%%%%%%%%%%%%%%%%%%%%%%%%%%%%%%%%%%%%%%%%%%%%%%%%%%%%%%%%%%%%%%%%%
%%				Appendix												%%
%%%%%%%%%%%%%%%%%%%%%%%%%%%%%%%%%%%%%%%%%%%%%%%%%%%%%%%%%%%%%%%%%%%%%%%%

\begin{widetext}
\section{Supplemental material to}
\begin{center}
{\large\bf "Superconducting quantum spin-Hall systems with giant orbital g-factors"}

\vskip 0.25cm

Rolf W. Reinthaler, Grigory Tkachov, and Ewelina M. Hankiewicz

\vskip 0.25cm

\textit{Institute for Theoretical Physics and Astrophysics, W\"urzburg University, Am Hubland, 97074 W\"urzburg, Germany}
\end{center}

\label{sec:Appendix}

\subsection{Details of the model}
The system is described by the Hamiltonian $\HBdG$ given by Eqs. (\ref{eq:Ham:BdG}) and (\ref{eq:Ham:3DTI}). 
We introduce three independent sets of Pauli matrices, $\pi_i$, $\tau_i $, and $\sigma_i$, to represent  
the particle-hole, top-bottom, and spin degrees of freedom, respectively, and write $\HBdG$ as
\begin{align}
	\label{eqApp:Hamiltonian}
	\HBdG =\hbar v k_x \pO \tz \sx + v e A_x \pz \tz \sx +  \hbar v k_y \pz \tz \sy + v e A_y \pO \tz \sy - \mu \pz \tO\sO 
					+ \Gamma \pz \tx \sO - \Delta  \py \tO \sy.
\end{align}
In this subsection we determine the vector potential $A_x$, taking into account the screening of the external field by the diamagnetic current in the superconductor (SC). 
We will also show that the vector potential acting on the states in the quantum spin-Hall system (QSHS) can be approximated by the value of $A_x$ at the SC surface.

\begin{figure}
	\begin{center}
		\includegraphics[width=0.3\linewidth]{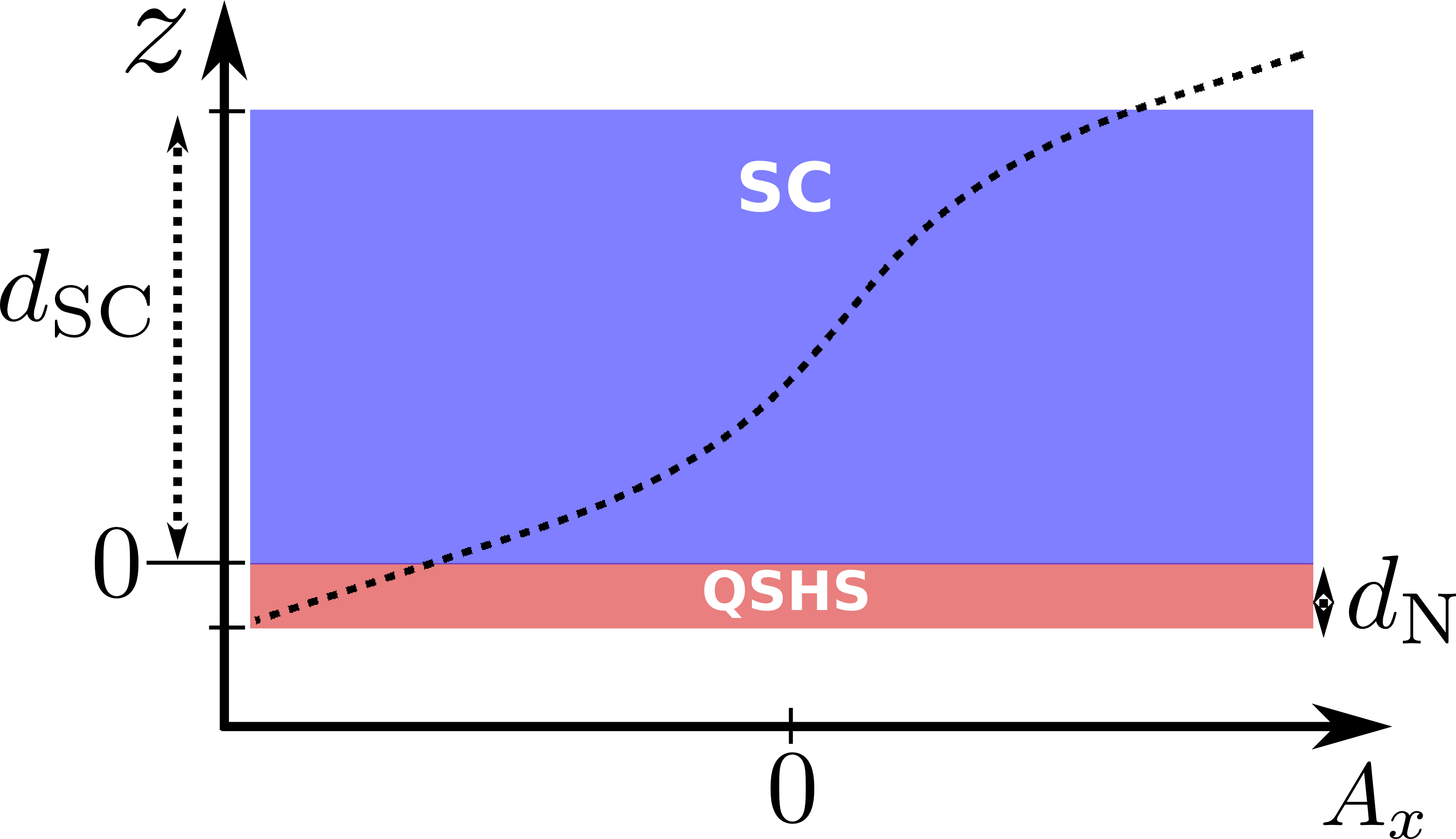}
		\caption{Geometry of SC/QSHS hybrid structure. The dashed curve shows the position dependence of the vector potential in Eq. (\ref{eqApp:Ax:Sol}).}
		\label{fig:SC:profile}
	\end{center}
\end{figure}

We choose the $z=0$ plane at the SC/QSHS interface (see also Fig. \ref{fig:SC:profile}) and assume that an external magnetic field is applied along the $y$ axis. 
It is convenient to use the London gauge
\begin{align}
	\label{eq:Vectorpot}
		\vec{A} = A_x (z) \hat{x},
\end{align}
in which the SC order parameter can be chosen real, so that the screening current density can be described by the London equation 
\begin{align}
	\vec{j} = -\frac{n_s e^2}{m^*} \vec{A},
\end{align}
where $m^*$ is the effective electron mass, and $n_s$ is related to the density of the Cooper pairs in the SC. 
Then, Amp\`ere's law in our geometry yields
\begin{align}
	\partial_z^2 A_x(z) =\mu_0 \frac{n_s e^2}{m^*} A_x(z)=\frac{1}{\lambda_L^2} A_x(z),
\label{Ampere}
\end{align}
with the London length $\lambda_L = \sqrt{m^*/(\mu_0 n_s e^2)}$. The boundary conditions to Eq. (\ref{Ampere}) are symmetric with respect to the applied field:
\begin{align}
	\partial_z A_x(z=0) = B_y, \qquad \partial_z A_x(z =\dsc) = B_y.
\label{Ampere_BC}
\end{align} 
Here $\dsc$ is the thickness of the superconductor in $z$-direction. 
Consequently, the solution of this boundary problem is anti-symmetric with respect to the middle of the SC:
\begin{align}
\label{eqApp:Ax:Sol}
	A_x(z) = B_y\lambda_L\frac{\sinh\left[\frac{1}{\lambda_L} \left(z - \frac{\dsc}{2}\right)\right]}{\cosh\left[\frac{\dsc}{2\lambda_L}\right]}, \qquad  0\leq z \leq \dsc.
\end{align}
By continuity, the vector potential in the QSHS is 
\begin{align}
\label{eqApp:Ax:QSHS}
	A_x(z) = B_y z + A_x(+0) = B_y z - B_y \lambda_L \tanh \frac{\dsc}{2\lambda_L}, \qquad -d_\mathrm{N} \leq z \leq 0,
\end{align}
where the first term is the vector potential of the external field, while $A_x(+0)$ is the solution (\ref{eqApp:Ax:Sol}) at the surface $z=0$. 
Since for typical SC/QSHS structures the thickness of the QSHS $d_\mathrm{N} \ll \lambda_L \tanh\left(\dsc/2\lambda_L\right)$, we can approximate 
\begin{align}
\label{eqApp:Ax:QSHS1}
	A_x(z) \approx A_x(+0) = - B_y \lambda_L \tanh \frac{\dsc}{2\lambda_L}, \qquad -d_\mathrm{N} \leq z \leq 0.
\end{align}
In the main text we often compare the energy scale associated with the magnetic field, $evA_x$, with the superconducting gap $\Delta$. 
Hence it is useful to introduce the dimensionless parameter  
\begin{align}
	\frac{|e v A_x|}{\Delta} = \frac{|B_y|}{B_*}, \quad\text{with } B_* = \frac{\Delta}{e v \lambda_L \tanh\left(\dsc/2\lambda_L\right)} = 
\frac{\phi_0}{\pi \xi_{\mathrm{SC}} \lambda_L \tanh\left(\dsc/2\lambda_L\right)},
\end{align}
where we introduced the characteristic magnetic field $B_*$ at which the gap closes as well as the SC coherence length and the flux quantum
\begin{align}
	\xi_{\mathrm{SC}} = \frac{\hbar v}{\Delta}, \quad \phi_0 =\frac{h}{2e}.
\end{align} 

\subsection{Solutions}
For solving $\HBdG$ in the semi-infinite plane $y>0$ using the boundary conditions 
\begin{align}
	\label{eqApp:BC}
 \psi(y=0) = \pO \ty \sy \psi(y=0).
\end{align}
 we use the ansatz
\begin{align}
	\psi = \Phi_k \exp\left[\imag k x - \frac{y}{\delta}\right].
\end{align}
Plugging it into the stationary Schr\"odinger equation we arrive at
\begin{align}
	\left[v\hbar k \pO \tz \sx + v e A_x \pz \tz \sx + \frac{\imag \hbar v}{\delta} \pz \tz \sy - \mu \pz \tO \sO + \Gamma \pz \tx \sO - \Delta\py\tO\sy %\Delta \cos\phi \py \tO \sy - \Delta \sin\phi \px \tO \sy 
	\right] \Phi_k = E \Phi_k.
\end{align}
At $y=0$, where \eqref{eqApp:BC} holds, we use 
\begin{align}
	 \pz\tx \sO \Phi_k = \pz \tx \sO (\pO\ty\sy)\Phi_k = \pz (\imag \tz) \sy \Phi_k
\end{align}
to cast the sum $(\imag \hbar v/\delta) \pz \tz \sy + \Gamma \pz \tx \sO$ into $(\Gamma + \hbar v/\delta) \pz \tx \sO$. 
Equating this term to zero, we find the decay length 
\begin{align}
	\delta = - \frac{\hbar v}{\Gamma} =\frac{\hbar v}{|\Gamma|}.
\end{align}
The edge solution exists only for the inverted band structure with $\Gamma<0$. 
From the boundary conditions at $y=0$ we additionally find that the spinor simplifies to
\begin{align}
	\Phi_k = \left(\chi_k , \imag \sy \chi_k, \xi_k , \imag \sy \xi_k \right)^T.
\end{align}
Now we can reformulate the problem as a system of coupled linear equations
\begin{subequations}
 \label{eqApp:eqsys:1int}
 \begin{align}
  0 &= \left[v (\hbar k + e A_x)\sx - (\mu +E)\sO \right]\chi_k + \Delta\imag %\dphase 
  \sy \xi_k  \label{eqApp:eqsys:1int:1}\\
  0 &= \left[ -v (\hbar k + e A_x )\sx - (\mu + E ) \sO\right] \imag \sy \chi_k+\Delta\imag %\dphase 
  \sy(\imag\sy) \xi_k \label{eqApp:eqsys:1int:2}\\
  0 &= - \Delta \imag %\dphaseC 
  \sy \chi_k  + \left[v (\hbar k - e A_x)\sx + (\mu -E)\sO \right] \xi_k \label{eqApp:eqsys:1int:3}\\
   0 &= - \Delta \imag %\dphaseC 
   \sy (\imag \sy)\chi_k  + \left[-v (\hbar k - e A_x)\sx + (\mu -E)\sO \right] \imag \sy\xi_k. \label{eqApp:eqsys:1int:4}
 \end{align}
\end{subequations}
Using 
\begin{align}
	\sigma_i\sigma_j = \delta_{ij} \sO + \imag \epsilon_{ijk} \sigma_k
\end{align}
it is easy to show that upon multiplication by $-\imag \sy$ equations \eqref{eqApp:eqsys:1int:2} and \eqref{eqApp:eqsys:1int:4} correspond to equations \eqref{eqApp:eqsys:1int:1} and \eqref{eqApp:eqsys:1int:3}, respectively. We hence solve equations \eqref{eqApp:eqsys:1int:1} and \eqref{eqApp:eqsys:1int:3} independently. From \eqref{eqApp:eqsys:1int:1} we find
\begin{align}
	\label{eqApp:1int:spinor:xik}
	\xi_k = \frac{%\dphaseC
	1}{\Delta} \left[v (\hbar k + e A_x ) \sz - \imag (\mu +E) \sy \right]\chi_k.
\end{align}
Plugging this into \eqref{eqApp:eqsys:1int:3} we get
\begin{align}
	0 &= \left[\left( 2 v \hbar k \mu - 2 v e A_x E\right)\sz - \left(v^2 (\hbar^2 k^2 - e^2 A_x^2 ) + \mu^2 - E^2 + \Delta^2 \right) \imag \sy \right] \chi_k\quad \vert \cdot \imag \sy\nonumber\\
	0 &= -\left( 2 v\hbar k \mu - 2 v e A_x E\right) \sx \chi_k  + \left(v^2 (\hbar^2 k^2 - e^2 A_x^2 ) + \mu^2 - E^2 + \Delta^2 \right) \sO \chi_k
\end{align}
so that $\chi_k $ must be an eigenstate to $\sx$ having the eigenvalues $s = \pm1 $ with corresponding eigenvectors $(s,1)^T$. Finally we can solve the energy spectrum at the interface
\begin{align}
	\label{eqApp:1int:disp}
	E_{\pm}^s =  s ve A_x + \alpha \sqrt{\Delta^2 + (\mu - s v \hbar k)^2} = s ve A_x + \alpha W_{ks},
\end{align}
where $\alpha = \pm 1$ and designates conduction or valence bands. Further,  we defined 
\begin{align}
	W_{ks} =\sqrt{\Delta^2 + (\mu - s v \hbar k)^2}.
\end{align}
For the states we obtain
\begin{align}
	\chi_{ks} &= (s,1)^T, \quad \imag \sigma_y \chi_{ks} = (1 ,-s)^T \nonumber \\
	\xi^\pm_{ks} &= \frac{%\dphaseC
	1}{\Delta} \left[ \hbar v k \sigma_z + (-\mu  \mp W_{ks})\imag \sigma_y \right] \chi_{ks}\nonumber \\
	\imag \sigma_y \xi^\pm_{ks} &= \frac{%\dphaseC
	1}{\Delta} \left[ -\hbar v k \sigma_x + (\mu  \pm W_{ks})\imag \sigma_0 \right] \chi_{ks}
\end{align}
\begin{align}
	\label{eqApp:1int:spinor}
	\psi^{\pm}_{k s} 
			&=  \frac{1}{\mathcal N}\exp\left[\imag k x - y \frac{\vert \Gamma\vert}{\hbar v}\right] \begin{pmatrix}
																						\mu^\pm_{ks}\phi_s\\
																						\nu^\pm_{ks} \phi_{-s}\\
																					\end{pmatrix},
\end{align}
where $\mathcal{N}$ comes from the normalization of the wave function, 
\begin{align}
 \phi_s = \left(s, 1, 1, -s \right)^T
\end{align}
is the particle spinor and 
\begin{align}
	\mu^\pm_{ks} =1 , \quad \nu^\pm_{ks}=\frac{%\dphaseC
	1}{\Delta}\left(-  v \hbar k +s \mu \pm s \sqrt{\Delta^2 + (\mu -s v\hbar k)^2}\right)
\end{align}
are the electron and hole weights, respectively. In this way we have decoupled the solution in a part coming from the particle solution $\exp\left[\imag k x - y \frac{\vert \Gamma\vert}{\hbar v}\right] \phi_s$ and a part containing the superconducting order parameter. \\
The normalization consists of two parts. The first is the normalization of the particle and hole weights
\begin{align}
\label{eqApp:1int:normalization:spinor}
	\tilde{N}^2 = \vert \mu^\pm_{ks} \vert^2 +   \vert \nu^\pm_{ks} \vert^2= \frac{2}{\Delta^2} \left[W_{ks}^2 \pm (\mu - s v \hbar k ) W_{ks}\right].
\end{align}
Additionally we have to normalize over the half space $y>0$ and the dimensions of the spinor 
\begin{align}
	\label{eqApp:1int:normalization:2}
	N^2 =  \int_0^{\infty} \rd y \exp\left[-2 y \frac{\vert \Gamma\vert}{\hbar v}\right] \phi_s^*\cdot \phi_s = (2 +2s^2) \frac{\hbar v}{2 \vert \Gamma\vert} = 2 \frac{\hbar v}{ \vert \Gamma\vert}.
\end{align}
The full normalization is given by
\begin{align}
	\label{eqApp:1int:normalization}
		\mathcal N = \tilde N N = \frac{2}{\Delta} \sqrt{W_{ks}^2 \pm (\mu - s \hbar k v) W_{ks}} \sqrt{\frac{\hbar v}{\vert \Gamma\vert}}.
\end{align}
It is advantageous to rewrite the spinor \eqref{eqApp:1int:spinor} in the language of the coherence factors, because this form reflects the natural symmetries of the system. 
We define the dimensionless parameter 
\begin{align}
	\label{eqApp:eta:ks}
	\eta_{ks} = \frac{\mu - sv \hbar k }{\Delta},
\end{align}
which measures the deviation away from the Fermi points. With this $W_{ks}= \sqrt{\Delta^2 + \eta_{ks}^2}/\vert\Delta\vert$. 
The coherence factors are defined by ($\Delta>0 \Leftrightarrow \sign\Delta =1 \Leftrightarrow \vert \Delta \vert = \Delta$)
\begin{align}
	\label{eqApp:1int:CoherenceFactor:u:ks}
	u^\alpha_{ks} &= \frac{1}{N(\alpha, k, s)} = \frac{1}{\sqrt{2}}\frac{\vert \Delta\vert }{\sqrt{W_ks^2 + \alpha (\mu - s v \hbar k) W_{ks}}} \nonumber \\
	&= \frac{1}{\sqrt{2}} \frac{1}{\sqrt{1 + \eta_{ks}^2 + \alpha \eta_{ks} \sqrt{1 + \eta_{ks}^2}} } \\
	\label{eqApp:1int:CoherenceFactor:v:ks}
	v^\alpha_{ks} &= \frac{\nu_{ks}^\alpha}{N(\alpha, k, s)}= \frac{%\dphaseC
	1}{\sqrt{2}} \frac{ s \;\sign \Delta (\mu - s v \hbar k +\alpha W_{ks} )}{\sqrt{W_ks^2 + \alpha (\mu - s v \hbar k) W_{ks}}} \nonumber \\
	 &= s \frac{%\dphaseC
	 1}{\sqrt{2}} \frac{ \eta_{ks} + \alpha \sqrt{1 + \eta_{ks}^2}}{\sqrt{1 + \eta_{ks}^2 + \alpha \eta_{ks} \sqrt{1 + \eta_{ks}^2}}}
	 =  \frac{%\dphaseC
	 1}{\sqrt{2}} \frac{s\; \sign\left[\eta_{ks} + \alpha \sqrt{1 + \eta_{ks}^2}\right]}{\sqrt{\frac{1 + \eta_{ks}^2 + \alpha \eta_{ks} \sqrt{1 + \eta_{ks}^2}}{(\eta_{ks} + \alpha \sqrt{1 + \eta_{ks}^2})^2}}}\nonumber\\
	 &=  \frac{%\dphaseC
	 1}{\sqrt{2}} \frac{s\; \sign\left[\eta_{ks} + \alpha \sqrt{1 + \eta_{ks}^2}\right]}{\sqrt{1 + \eta_{ks}^2 - \alpha \eta_{ks} \sqrt{1 + \eta_{ks}^2}} }= \frac{%\dphaseC
	 1}{\sqrt{2}} \frac{s \alpha}{\sqrt{1 + \eta_{ks}^2 - \alpha \eta_{ks} \sqrt{1 + \eta_{ks}^2}} }.
\end{align}
In the last line we used 
\begin{align}
	1 + \eta_{ks}^2 + \alpha \eta_{ks} \sqrt{1 + \eta_{ks}^2 } =( 1 + \eta_{ks}^2 - \alpha \eta_{ks} \sqrt{1 + \eta_{ks}^2 } )(\eta_{ks} + \alpha \sqrt{1 + \eta_{ks}^2})^2
\end{align}
as well as that for propagating solutions (real $k$): $\eta_{ks} \in \mathbb{R} \Rightarrow \eta_{ks}^2 >0 \Rightarrow \sqrt{1 +\eta_{ks}^2} > \vert\eta_{ks}\vert$. Therefore we can simplify the $\sign$ expression to $\sign \alpha = \alpha$.\\
Since we will be mainly interested in energy dependent quantities, like the elastic scattering and zero energy states, it is useful to reformulate the coherence factor in an energy dependent form. We use 
\begin{align}
	\label{eqApp:1int:kvec}
 	k^\beta_s(E) = \frac{1}{v\hbar} \left(s \mu +\beta \sqrt{(E - s e v A_x)^2 - \Delta^2 } \right), \quad \beta = \pm1.
\end{align}
In squaring the dispersion \eqref{eqApp:1int:disp} we loose the information $\alpha=\pm$, so that $\beta$ is not in one to one correspondence with $\alpha$. The relation can be found by checking the consistency of equation 
\begin{align}
	E^\alpha_s =  s v e A_x + \alpha \sqrt{\Delta^2 + (\mu - s v \hbar k^\beta_s(E))^2} \stackrel{!}{=}E,
\end{align} 
where we fix $E$ and $\beta$. The result is independent of $\beta$: 
\begin{align}
	\label{eqApp:1int:alphachoice}
	E^\alpha_s =  s v e A_x + \alpha \vert E - s e v A_x\vert \quad \Rightarrow \quad\alpha = 	\begin{cases}
																								1, &E > s e v A_x\\
																								-1, &E<  s e v A_x
																							\end{cases}
\end{align}
and $\alpha$ is fixed by energetic constraints. In total the energy dependent formulation takes the form
\begin{align}
	\label{eqApp:1int:CoherenceFactor:u:Es}
	u^\beta_{Es} &= \frac{1}{\sqrt{2}} \frac{\Delta}{\sqrt{(E - s e v A_x)^2 - s \beta (E - s e v A_x) \sqrt{(E - s e v A_x)^2 -\Delta^2} }}\\
	\label{eqApp:1int:CoherenceFactor:v:Es}
	v^\beta_{Es} &= \frac{%\dphaseC
	1}{\sqrt{2}} \frac{s \Delta \sign \left[E - s e v A_x\right] }{\sqrt{(E - s e v A_x)^2 + s \beta (E - s e v A_x) \sqrt{(E - s e v A_x)^2 -\Delta^2} }}.
\end{align}
The full wave functions are 
\begin{align}
	\label{eqApp:1int:states:CoherenceFactors}
	\psi_{ks}^\alpha(y) = \frac{\exp\left[ \imag kx - y \frac{\vert \Gamma \vert}{\hbar v}\right]}{N_2} \begin{pmatrix}  u_{ks}^\alpha \phi_s \\ v_{ks}^\alpha \phi_{-s}\end{pmatrix}, \quad \psi_{Es}^\beta(y) = \frac{\exp\left[ \imag k^\beta_{Es} x - y \frac{\vert \Gamma \vert}{\hbar v}\right]}{N_2} \begin{pmatrix}  u_{Es}^\beta \phi_s \\ v_{Es}^\beta \phi_{-s}\end{pmatrix}.
\end{align}
Let us analyze the coherence factors further. For both representations it is easy to show that 
\begin{align}
	\vert u_{ks}^\alpha \vert^2 + \vert v_{ks}^\alpha \vert^2 = \vert u_{Es}^\beta \vert^2 + \vert v_{Es}^\beta \vert^2 = 1.
\end{align}
Further it is interesting to note that
\begin{align}
	\label{eqApp:1int:CoherenceFactor:Relation}
	v^\beta_{Es} = s\; \sign\left[E - s e vA_x\right] %\dphaseC 
	u_{Es}^{-\beta}, \quad v^\alpha_{ks} =\alpha %\dphase 
	u^{-\alpha}_{ks}, 
\end{align}
i.e. the coherence factors are, up to a phase factor, related by $\beta \rightarrow -\beta$ ($\alpha \rightarrow -\alpha$).  

The quasi-particle character ( the particle or the hole character of the excitations) of a state can be calculated using the operator $\pz \tO \sO$
\begin{align}
	P^\beta_{Es} &= \left(u^{\beta}_{Es}\right)^* u^{\beta}_{Es}   - \left(v^{\beta}_{Es}\right)^* v^{\beta}_{Es} = s \beta \frac{\sqrt{(E-s e v A_x)^2 -\Delta^2}}{E- s e v A_x} \\
	P^\alpha_{ks} &=\left(u^{\alpha}_{ks}\right)^* u^{\alpha}_{ks}   - \left(v^{\alpha}_{ks}\right)^* v^{\beta}_{Es}=- \alpha \frac{\mu - s \hbar v k}{\sqrt{\Delta^2 + (\mu - s \hbar v k)^2}},
\end{align}
showing how the quasi-particle character depends on  $\alpha$ ($\beta$), $s$ and $k$ ($E$). For a given energy and spin, $\beta$ indicates the quasi-particle character of the state.
 At the Fermi points $k^F= \frac{s \mu}{\hbar v}$ ($E^F = E_{k^Fs}^{\alpha} = s v e A_x + \alpha \Delta$) one finds
\begin{align}
	P^\beta_{E^Fs} = P^\alpha_{k^Fs} =0,
\end{align}
i.e. a perfect mixture of electrons and holes at the band edges. Far away from the quasi-particle gap we get ($k =\kappa \vert k\vert, \; E = \epsilon \vert E \vert$, where $\kappa = \epsilon = \pm1$)
\begin{align}
\lim_{\vert E\vert\rightarrow  \infty} P^\beta_{Es} = \epsilon s \beta, \quad \lim_{\vert k\vert\rightarrow  \infty} P^\alpha_{ks} = \kappa s \alpha,
\end{align}
where $P\rightarrow 1$ ($P\rightarrow -1$) indicates electrons (holes).

	\subsection{DOS and sDOS}
The spin-dependent density of states in the energy interval $[ E , E+\rd E]$ at position $y$ is defined by
\begin{align}
	\rho_s (y, E) = -\frac{1}{\pi} \mathrm{Tr} \;\left[\Im G^R_s (y,y)\right]_{11},
\end{align}
where the retarded Green's function is given in Lehmann representation
\begin{align}
	\label{eqApp:GreenFKT}
	G^R_s (y, y') &= \sum_{k, \alpha=\pm} \frac{\psi^\alpha_{ks} (y) \otimes \left( \psi^\alpha_{ks} (y')\right)^\dagger}{E - E^s_{\alpha} (k) + \imag 0^+} \\
	&=  \sum_{k, \alpha=\pm} \frac{\psi^\alpha_{ks} (y) \otimes \left( \psi^\alpha_{ks} (y')\right)^\dagger}{(E - E^s_{\alpha} (k))^2 + ( 0^+)^2}\left[E - E^s_{\alpha} (k) - \imag 0^+ \right].
\end{align}
The $11$ component $\left[\Im G^R_s (y,y')\right]_{11}$ corresponds to the electron block of the Green's function. 
The diagonal parts of $\psi^\alpha_{ks} (y) \otimes \left( \psi^\alpha_{ks} (y')\right)^\dagger$ are purely real, because the $y$-dependence is an exponential decay instead of a plain wave. Using the Lorentz representation of the Dirac delta-distribution
\begin{align}
	\delta(x) = \lim\limits_{\epsilon \rightarrow 0^+} \delta_\epsilon(x) = \lim\limits_{\epsilon \rightarrow 0^+} \frac{1}{\pi}\frac{\epsilon}{x^2 + \epsilon^2}
\end{align}
we find that the diagonal parts of the imaginary part of the Green's function can be written as
\begin{align}
\Im \left(G^R_s(y,y')\right)_{m,m} &= - \sum_{k, \alpha = \pm} 0^+ \frac{\left(\psi^\alpha_{ks} (y) \otimes \left( \psi^\alpha_{ks} (y')\right)^\dagger\right)_{m,m}}{(E - E^s_{\alpha} (k))^2 + ( 0^+)^2} \nonumber \\
&= -\pi \sum_{k, \alpha = \pm}\left(\psi^\alpha_{ks} (y) \otimes \left( \psi^\alpha_{ks} (y')\right)^\dagger \right)_{m,m}\delta \left(E - E_\alpha^s(k) \right).
\end{align}
Evaluating the trace we arrive at the simple form
\begin{align}
	\label{eqApp:sDOS}
	\rho_s(y, E) &= \sum_{k, \alpha = \pm} \left\Vert \left[\psi^{\alpha}_{ks} (E) (y)\right]_{e^-}\right\Vert^2 \delta \left(E - E_\alpha^s(k) \right) \nonumber\\
	&=\sum_{ k ,\alpha = \pm}  \sum_{\{k_0\vert E_\alpha^s(k_0) = E\}}\left\Vert \left[\psi^{\alpha}_{ks} (E) (y)\right]_{e^-}\right\Vert^2 \delta \left(k - k_0\right)  \left\vert\frac{\rd E_\alpha^s(k') }{\rd k'} \right\vert^{-1}_{k'=k_0}.
\end{align}
Here 
\begin{align}
	\label{eqApp:1int:Spinor:electron}
	\left[\psi^{\alpha}_{ks} (E) (y)\right]_{e^-}
			&= \frac{1}{N}\exp\left[\imag k x - y \frac{\vert \Gamma\vert}{\hbar v}\right]
	\begin{pmatrix} \chi_{ks} \\ \imag \sigma_y\chi_{ks} \end{pmatrix}	
\end{align}
is the electron part of the spinor \eqref{eqApp:1int:spinor}. In effect it is the summation over the propagating modes at energy $E$ corresponding to the quantum number $s$ weighted with the slope of the band. The total DOS at position $y$ is then given by the sum of $\rho_\up(y,E)$ and $\rho_\down(y,E)$. The full signal can further be obtained by integration over the half space $y>0$.\\	
In \eqref{eqApp:sDOS}, $\alpha$ is chosen by the energetic constraint Eq.~\eqref{eqApp:1int:alphachoice}. Later on we will find that $\rho_s(y,E)$ is independent of $\alpha$.\\
The main contribution to the spin-dependent density comes from the derivative of the band structure
\begin{align}
	\frac{\partial E^\alpha_s}{\partial k} = \alpha s v \hbar \frac{s v \hbar k-\mu}{\sqrt{\Delta^2 + (\mu - s v \hbar k)^2}}
\end{align}
so that 
\begin{align}
	\label{eqApp:1int:InvD}
	\left\vert\frac{\rd E_\alpha^s(k') }{\rd k'} \right\vert^{-1}_{k'=k_s^\beta} = \frac{1}{v\hbar} \frac{\vert E - s e v A_x\vert}{\sqrt{(E- s e v A_x)^2 - \Delta^2}}.
\end{align}
It shows poles at the band edges
\begin{align}
	\label{eqApp:1int:bandedges}
	E =  s e v A_x \mp \Delta.
\end{align}
and goes to $(\hbar v)^{-1}$ away from the superconducting gap, when $E- s e v A_x \gg \Delta$. The density is renormalized by taking only the $y$-dependent electron contribution ($D_{Es} = E- sev A_x$)
\begin{align}
	\left\Vert\left[\psi^{\alpha}_{k^\beta_s (E)s} (E) (y)\right]_{e^-} \right\Vert^2 =& \frac{1}{N^2} \exp\left[-2 y \frac{\vert \Gamma\vert}{\hbar v}\right] (s^2 + 1+1+ s^2) \nonumber\\
	&=\exp\left[-2 y \frac{\vert \Gamma\vert}{\hbar v}\right] \frac{\vert \Gamma \vert }{\hbar v} \Delta^2 \frac{1}{D_{Es}^2 + \alpha \beta \sqrt{D^2_{Es} -\Delta^2 } \vert D_{Es}\vert } \nonumber \\
	&=\exp\left[-2 y \frac{\vert \Gamma\vert}{\hbar v}\right] \frac{\vert \Gamma \vert }{\hbar v} \Delta^2 \frac{1}{D_{Es}^2 +  \beta \sqrt{D^2_{Es} -\Delta^2 }D_{Es}}.
\end{align}
Again the $\alpha$ dependence is obsolete, since - accordingly to \eqref{eqApp:1int:alphachoice} - $\vert D_{Es}\vert =\vert E - s e v A_x\vert$ cancels the choice of $\alpha$. Hence
\begin{align}
	\rho_s(y, E) &= \sum_{\beta=\pm}  \left\Vert \left[\psi^1_{k^\beta_s(E)}  (y)\right]_{e^-}\right\Vert^2   \left\vert\frac{\rd E_1^s(k') }{\rd k'} \right\vert^{-1}_{k'=k^\beta_s(E)}\left(\Theta[ -D - \Delta ] + \Theta[D-\Delta]\right).
	%	\left(\Theta[+s e v A_x - \Delta -E ] + \Theta[E - (+s e v A_x + \Delta)]\right).
\end{align}
The Heavyside-$\Theta$ functions ensure that $E$ does not lie within the spin-dependent gap \eqref{eqApp:1int:bandedges}, which was previously enforced by the Dirac-$\delta$ distribution.
\\
Since \eqref{eqApp:1int:InvD} is independent of $\beta$, the summation simplifies to a factor of two.
In total we find the closed form
\begin{align}
	\label{eqApp:sDOS:Ana}
	\rho_s(y, E) =& \frac{2\vert \Gamma\vert}{\hbar v}\exp\left[-2 y \frac{\vert \Gamma\vert}{\hbar v}\right] \frac{1}{\hbar v} \frac{\vert E - s e v A_x\vert}{\sqrt{(E- s e v A_x)^2 - \Delta^2}}
	\left(\Theta[ s e v A_x - \Delta -E ] + \Theta[E - (s e v A_x + \Delta)]\right).
\end{align}
Integration over the half space gives
\begin{align}
	\label{eqApp:sDOS:Ana:INT}
	\rho_s(E) &= \int_0^\infty \rd y \rho_s(y,E) = \frac{1}{\hbar v} \frac{\vert E - s e v A_x\vert}{\sqrt{(E-s e v A_x)^2 - \Delta^2}}
	\left(\Theta[ s e v A_x - \Delta -E ] + \Theta[E - (s e v A_x + \Delta)]\right).
\end{align}
It is easy to see that the $s$-polarization $P_s = \rho_\up - \rho_\down$ vanishes for $A_x=0$, because all $s$ dependence vanishes. But also for $\Delta=0$ we find vanishing $P_s$, since
\begin{align}
	\left.\frac{1}{\hbar v} \frac{\vert E - s e v A_x\vert}{\sqrt{(E- s e v A_x)^2 - \Delta^2}}\right\vert_{\Delta=0} = \frac{1}{\hbar v} = \mathrm{const.}
\end{align}
together with $\Theta[E- s e v A_x] + \Theta[ s e v A_x-E] \equiv 1$, which is expected for an ungaped system. Again $\rho_\up$ and $\rho_\down$ are equivalent and $P_s$ is zero.
\\
Interestingly the spin-dependent density is independent of the chemical potential $\mu$. Comparing with the dispersion \eqref{eqApp:1int:disp} we find that $\mu$ just spin-dependently shifts $k$, while the density of states is calculated by summing over all $k$ values at a certain energy.
At zero vector potential the system shows to be a gaped superconductor, cf. Fig.~\ref{fig:1Int:GapClosing:a}. Increasing $A_x$ closes this gap. In Fig \ref{fig:1Int:GapClosing} we analyze the gap closing in more detail.
\begin{figure}[p]
	\begin{center}
		\subfloat[$E(k), \;B_y =0 B_*$]{\includegraphics[width=0.38\linewidth]{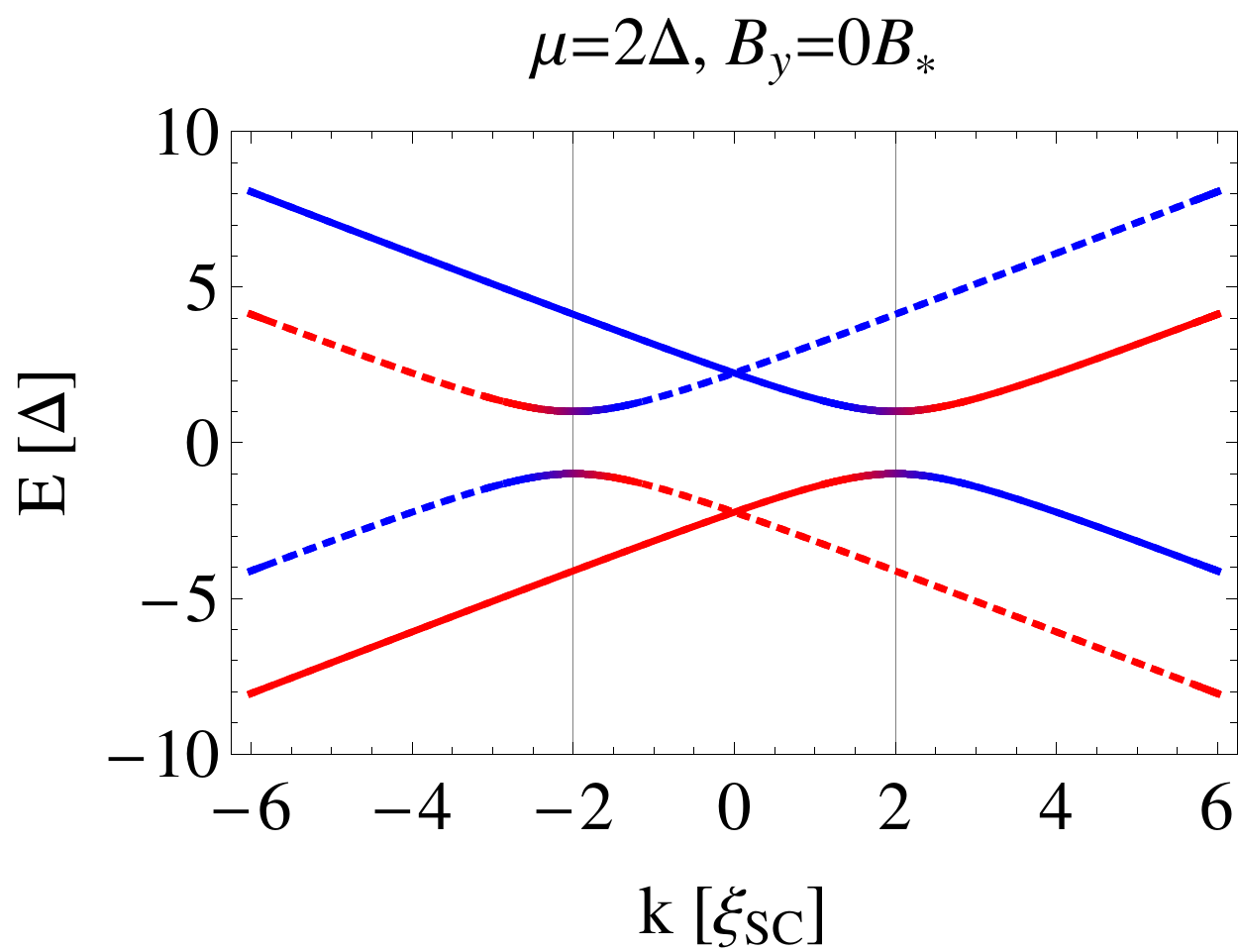}\label{fig:1Int:GapClosing:a}}\hfill
		\subfloat[$\rho(E), \;B_y =0 B_*$]{\includegraphics[width=0.36\linewidth]{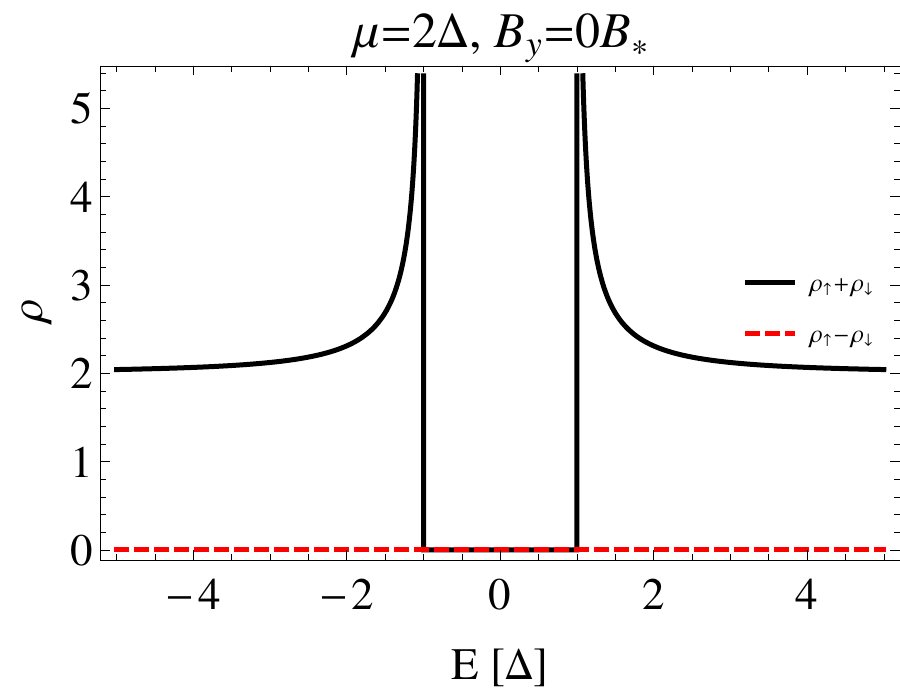}\label{fig:1Int:GapClosing:b}}\\
		\subfloat[$E(k), \;B_y =0.5 B_*$]{\includegraphics[width=0.38\linewidth]{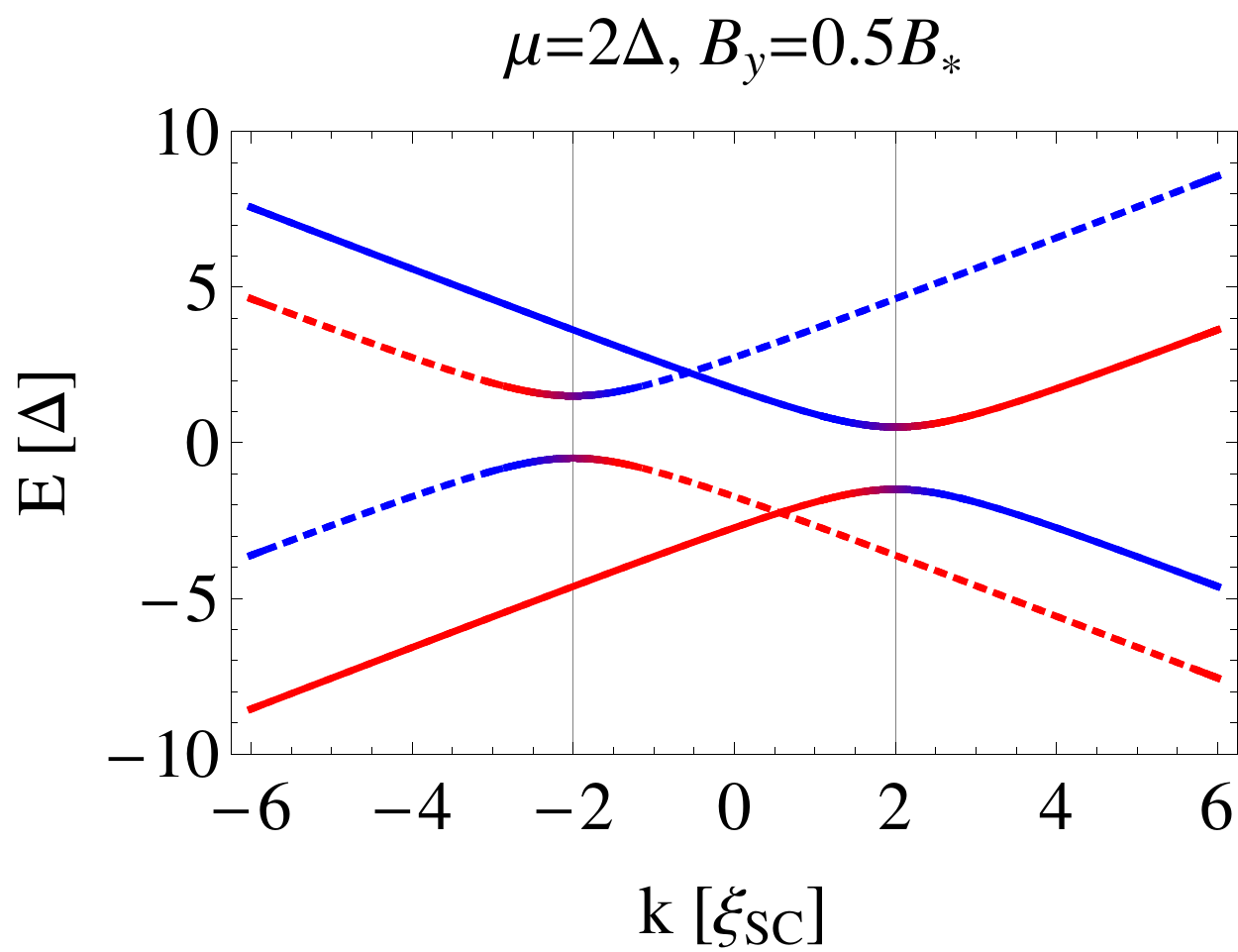}\label{fig:1Int:GapClosing:c}}\hfill
		\subfloat[$\rho(E), \;B_y =0.5 B_*$]{\includegraphics[width=0.36\linewidth]{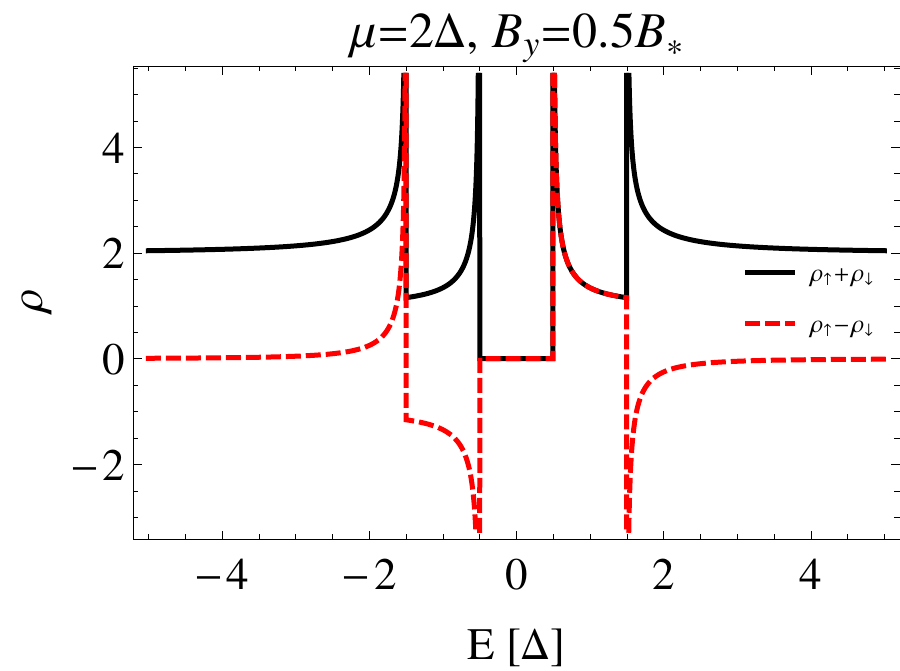}\label{fig:1Int:GapClosing:d}}\\
		\subfloat[$E(k), \;B_y =1 B_*$]{\includegraphics[width=0.38\linewidth]{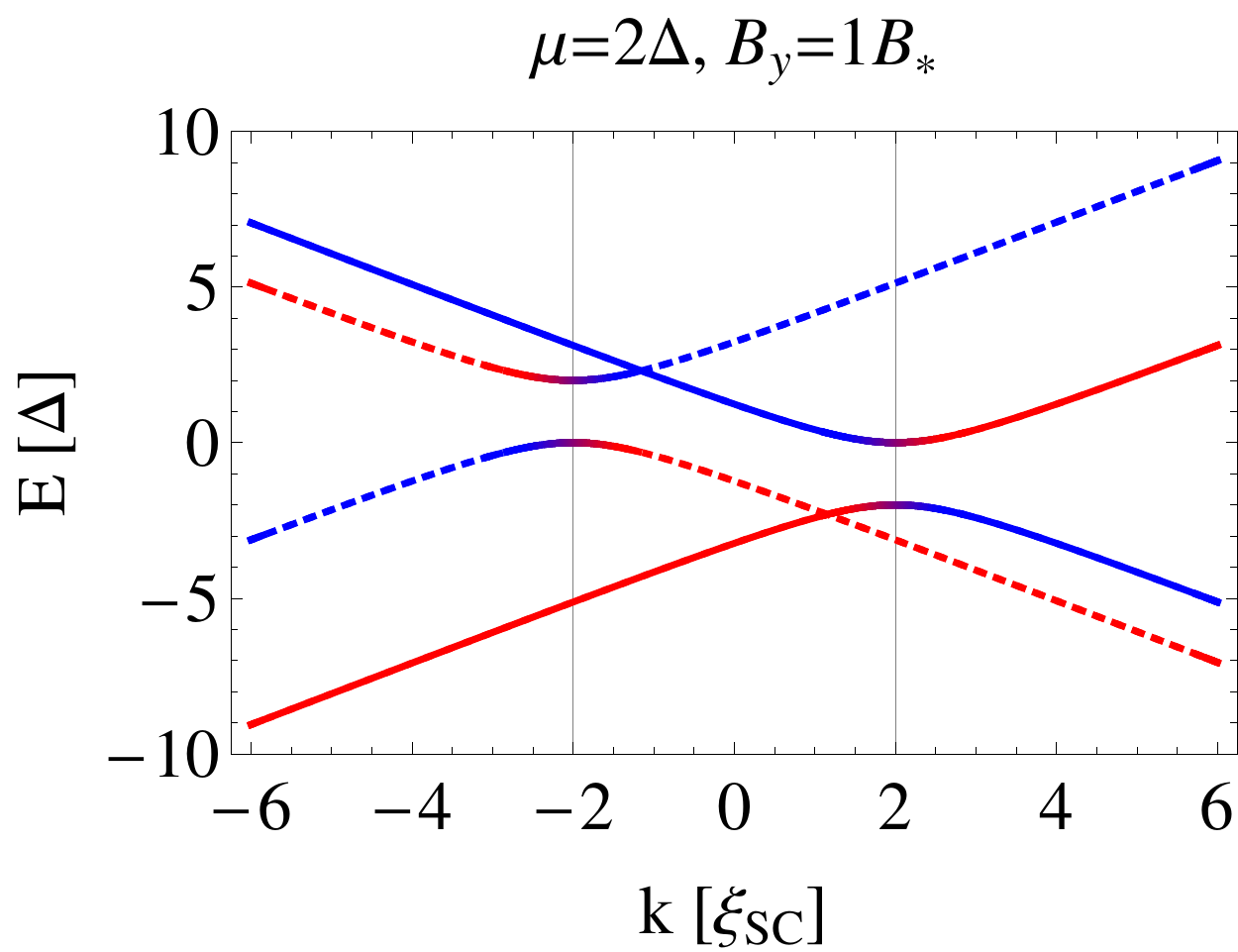}\label{fig:1Int:GapClosing:e}}\hfill
		\subfloat[$\rho(E), \;B_y =1 B_*$]{\includegraphics[width=0.36\linewidth]{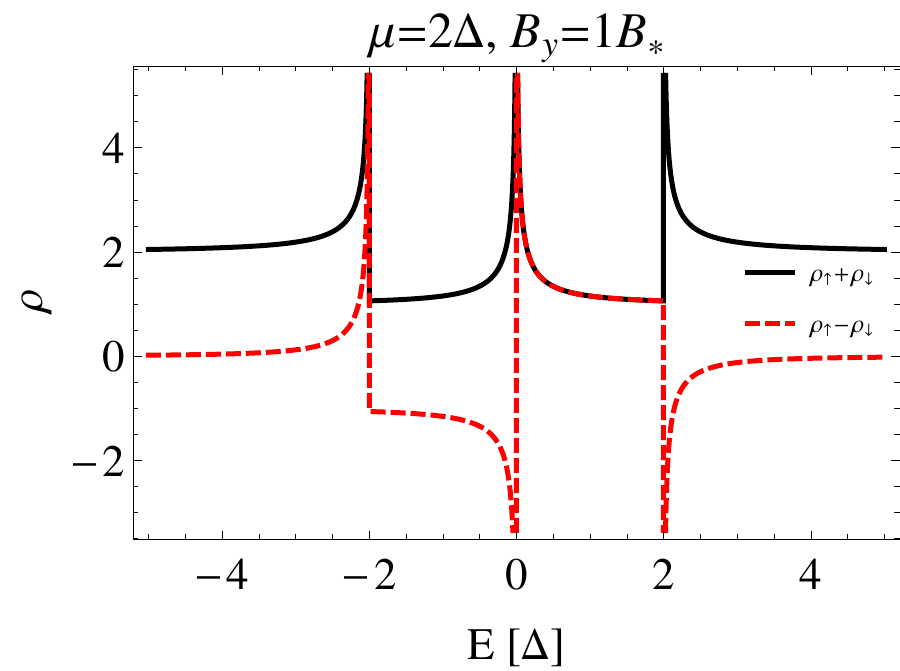}\label{fig:1Int:GapClosing:f}}\\
		\subfloat[$E(k), \;B_y =2.15 B_*$]{\includegraphics[width=0.38\linewidth]{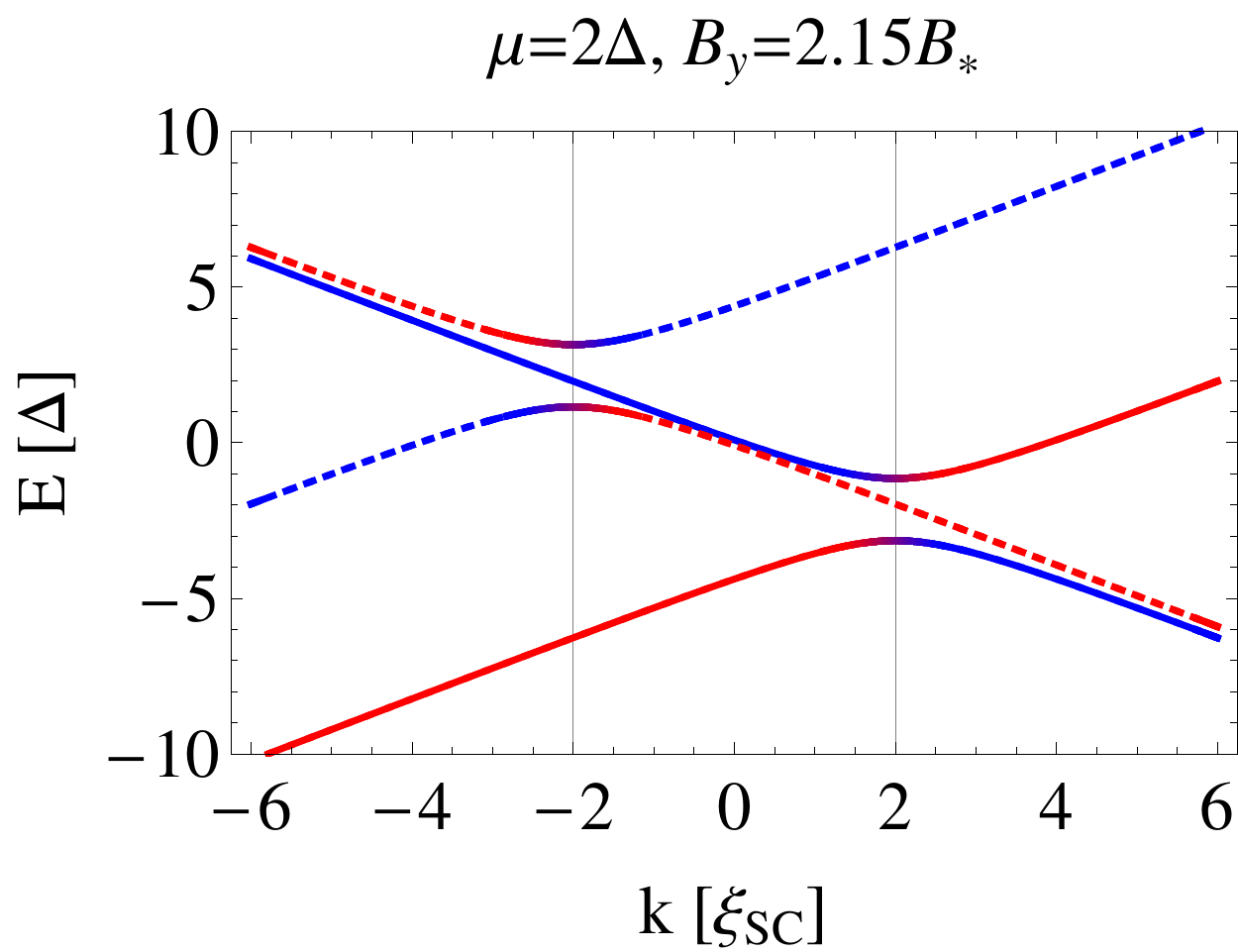}\label{fig:1Int:GapClosing:g}}\hfill
		\subfloat[$\rho(E), \;B_y =2.15 B_*$]{\includegraphics[width=0.36\linewidth]{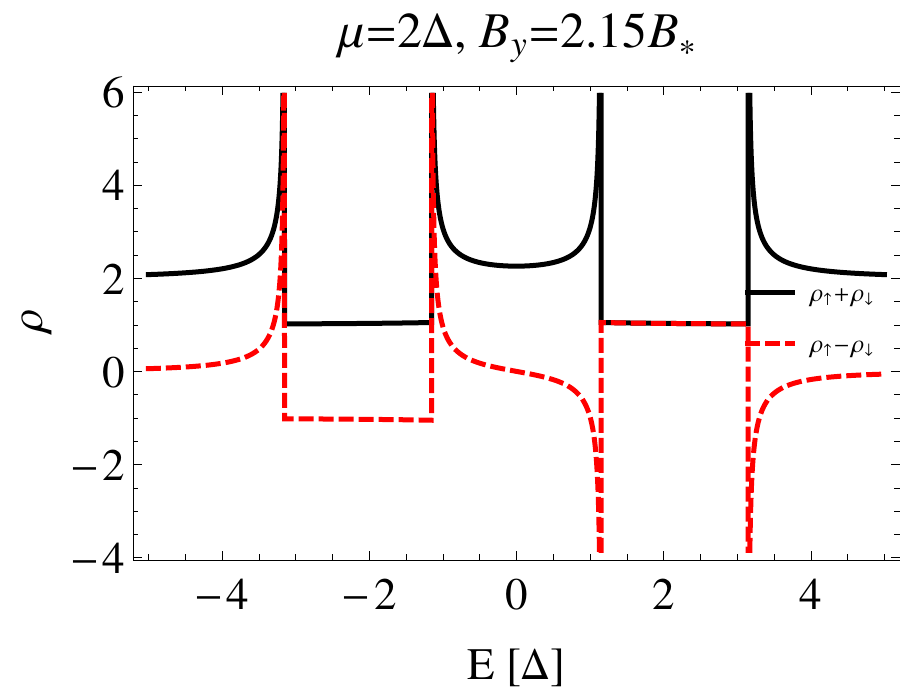}\label{fig:1Int:GapClosing:h}}
		\caption{The gap closing is analyzed by comparing the dispersion (left column) with the DOS (right column) for different values of $B_y$. In the left column $s=1$ ($-1$) is indicated by solid (dashed) curves. The particle character is color encoded, blue (red) being used for hole (electron) like states. }
		\label{fig:1Int:GapClosing}
	\end{center}
\end{figure}
For zero vector potential (cf. Figs.~\ref{fig:1Int:GapClosing:a} and \ref{fig:1Int:GapClosing:b}) we find that states of $\alpha=+ $ and $\alpha=-$ ( conduction and valence bands) are separated by a superconducting gap. Additionally we find a Dirac cone between states of same $\alpha$ but opposite $s$. The cone is protected because states of different $s$ are orthogonal. Increasing $A_x$ shifts the branches of different $s$ in opposite directions and therefore the gap opens around energies $E=s evA_x$, while  $k= \pm \mu/(\hbar v)$ is fixed. This is indicated by the vertical grid lines, cf. Fig.~\ref{fig:1Int:GapClosing:c}. Increasing the vector potential (see Fig.~\ref{fig:1Int:GapClosing:d})  is reflected by shifting part of the density into the superconducting gap. This is accompanied by a spin-dependent signal, since the states of different $s$ do no longer live at the same energy. \\
When $\vert evA_x \vert = \vert \Delta\vert$, i.e. $\vert B_y\vert=B_*$ in Figs.~\ref{fig:1Int:GapClosing:e} and \ref{fig:1Int:GapClosing:f}, the indirect gap in the dispersion is closed, indicated by a peak in the DOS at zero energy. However, the spin branches have a gap. 
Increasing the vector potential even further shifts the relative spin branches far enough that they switch energetic order at $B_y \approx  2.25B_*$ for our parameters, cf. Figs.\ref{fig:1Int:GapClosing:g} and \ref{fig:1Int:GapClosing:h}.

\subsection{Transport within the BTK formalism}
A way to measure the spin polarization of the edge states is transport through an interface between a QSHS and the QSHS in the proximity to the s-wave superconductor (NSC-junction).
A possible realization of such a junction is the Y-forked NSC junction described in the main text. 
Due to spin-momentum locking the incoming electronic state on the left side of the sample has a fixed spin on a certain edge of the sample. 
The normal part of the junction is described by $\HBdG$ with $\Delta\rightarrow 0$. Using $\sqrt{D_{Es}^2 + \Delta^2} \stackrel{\Delta \ll D_{Es}}{\approx} \vert D_{Es} \vert \left(1 - \frac{\Delta^2}{2 D_{Es}^2}\right)$ we find
\begin{align}
\label{eqApp:1int:states:Limit:delta}
	\lim_{\Delta\rightarrow 0}  u^\beta_{Es} &= \frac{1}{\sqrt{2}}\frac{\Delta}{\sqrt{D_{Es}^2 (1 - s\beta \sign[D_{Es}])  +  s \beta \sign[D_{Es}] \frac{\Delta^2}{2}}} = 
				\begin{cases}
					0, & \sign[D_{Es}] s \beta <0\\
					1, & \sign[D_{Es}] s \beta >0\\
				\end{cases}  \nonumber\\
	\lim_{\Delta\rightarrow 0}  v^\beta_{Es} &= \frac{s \sign[D_{Es}]}{\sqrt{2}}\frac{\Delta}{\sqrt{D_{Es}^2 (1+s\beta \sign[D_{Es}])  -  s \beta \sign[D_{Es}] \frac{\Delta^2}{2}}} \nonumber \\&= 
				\begin{cases}
					0, & \sign[D_{Es}] s \beta >0\\
					 s \sign[D_{Es}], & \sign[D_{Es}] s \beta <0\\
				\end{cases}.
\end{align}
The solutions in the normal regime hence take the form
\begin{align} 
	\text{electrons:}&  \quad \psi^{\beta( e^-)}_{Es} (y) &= \frac{\exp\left[\imag k^{\beta(e^-)}_{Es} x - y \frac{\vert \Gamma\vert }{\hbar v}\right]}{2} 
		\begin{pmatrix}
			\phi_s \\ 0\\
		\end{pmatrix} \nonumber \\
	\text{holes:}&  \quad \psi^{\beta( h^+)}_{Es} (y) &= \frac{\exp\left[\imag k^{\beta(h^+)}_{Es} x - y \frac{\vert \Gamma\vert }{\hbar v}\right]}{2} 
		\begin{pmatrix}
			0 \\ s \sign[D_{Es}]\phi_{-s}\\
		\end{pmatrix},	
\end{align}
where $\beta$ no longer is a degree of freedom but fixed by \eqref{eqApp:1int:states:Limit:delta}. For incoming electrons 
\begin{align}
	\beta(e^-) = s\; \sign[D_{Es}] 
\end{align}
has to be chosen. The helical nature of the edge states forbids electron reflection as long as the $\hat{\tau}\hat{C}$ symmetry is preserved. The only possible scattering channels are the electron transmission and reflection as a hole (Andreev reflection). When the excitation energy lies within the superconducting gap electron transmission must be zero and perfect Andreev reflection was predicted for both geometries, provided that the width of the ribbon is large enough to separate the counter propagating edge channels  \cite{Adroguer2010, Reinthaler2013}. From matching the wave functions at the interface at $x=0$ 
\begin{align}
	\begin{pmatrix} \phi_s \\0 \\  \end{pmatrix} + R_A^s \begin{pmatrix} 0\\ s \;\sign[D_{Es}] \phi_{-s} \\ \end{pmatrix} = \mathcal{T} \begin{pmatrix} u_{Es}^\beta \phi_s\\ v^\beta_{Es} \phi_{-s} \\ \end{pmatrix}
\end{align}
we find
\begin{align}
	\label{eqApp:NS:AndreevCoeff}
	R_A^s(E) = \dphase s \; \sign[D_{Es}] \frac{v^\beta_{Es} }{u^\beta_{Es}}.
\end{align}
where $R_A^s(E)$ and $\mathcal{T}$ are the Andreev reflection and electron transmission amplitudes, respectively.  
We investigate the spin-dependent currents within the BTK formalism \cite{Blonder1982}. When the voltage $V$ is applied across the junction, the current can be calculated by
\begin{align}
	\label{eqApp:BTK}
	I_{\mathrm{BTK}}^s = \frac{e}{h} \int\limits_{-\infty}^\infty \rd E (1 + \vert R_A^s\vert^2)\left(f(E- e V ) - f(E)\right).
\end{align}
Here $f(E)$ is the distribution function of the carriers at excitation energy $E$ and temperature $T$
\begin{align}
	f(E)  = \frac{1}{\exp\left[\frac{E}{k_B T}\right] +1}.
\end{align}
The excess current is defined by 
\begin{align}
	\label{eqApp:Iex}
	I_{\mathrm{ex}}^s = \frac{e}{h} \int\limits_{-\infty}^\infty \rd E 
	\vert R_A^s\vert^2\left(f(E- e V ) - f(E)\right)
\end{align}
It gives the current which is transported additionally to the ohmic one and is generally measured at very high voltages. In the main text we found that the spin-dependent excess current shows a maximum at finite $B_y$. To analyze this further let us have a look at the excess current at zero temperature
\begin{align}
	I_{\mathrm{ex}}^s \stackrel{T=0}{=} \frac{e}{h} \int\limits_0^{eV} \rd E \vert R_A^s \vert^2 = \frac{e}{h} \int\limits_{+s g_* \mu_{\mathrm{B}} B_y}^{e V +s g_* \mu_{\mathrm{B}} B_y} \rd D_{Es} \vert R_A^s\vert^2,
\end{align}
where in the last step we used that in the Andreev reflection coefficient $E$ and $B_y$ only appear via $D_{Es} = E +s g_* \mu_{\mathrm{B}} B_y$. $g_*$ is the effective g-factor defined in the main text and $\mu_{\mathrm{B}}$ the Bohr magneton. With
\begin{align}
	\label{eqApp:AndreevProb}
	\vert R_A^s \vert^2 =\left\vert  \frac{D_{Es} - s \beta(e^-) \sqrt{D_{Es}^2-\Delta^2}}{D_{Es} + s\beta(e^-) \sqrt{D_{Es}^2-\Delta^2}}\right\vert = \left\vert  \frac{D_{Es} - \sign D_{Es} \sqrt{D_{Es}^2-\Delta^2}}{D_{Es} +\sign D_{Es}  \sqrt{D_{Es}^2-\Delta^2}}\right\vert
\end{align}
 the maximum in $B_y$ can be found by setting
\begin{align}
	\partial_{B_y} I_{\mathrm{ex}} \propto\vert R_A^s (e V+s g_* \mu_{\mathrm{B}} B_y)\vert^2 -\vert R_A^s (+s g_* \mu_{\mathrm{B}} B_y)\vert^2 
	\stackrel{!}{=} 0.
\end{align}
With $\vert R_A^s \vert^2 (D_{Es}) = \vert R_A^s \vert^2(-D_{Es})$ we find that this equation can be solved by
\begin{align}
	 B_y^{\mathrm{max}}= - s \frac{eV}{2 g_* \mu_{\mathrm{B}}}.
\end{align}
Another typical way to measure non-local conductances are $\rd I/\rd V$ characteristics. From \eqref{eqApp:BTK} we obtain
\begin{align}
	\label{eqApp:dIdV}
	\frac{\rd I^s } {\rd V}  =  \frac{e}{h} \int\limits_{-\infty}^\infty \rd E (1 + \vert R_A^s\vert^2)\left(\frac{\partial f(E-eV)}{\partial V}\right).
\end{align}

\end{widetext}

\end{document}